\documentclass{ar-1col-AAM}

\usepackage{txfonts}

\firstpagenote{Published in the Annual Review of Nuclear and Particle Science 2022, 72:177–204.
	\newline
	https://doi.org/10.1146/annurev-nucl-110221-103625
	\newline
	\newline
	Copyright \copyright\ 2022 the author(s), CC BY 4.0, https://www.annualreviews.org}

    \newcommand{\ie}{\rm{i.\,e.\@}}
    \newcommand{\eg}{\rm{e.\,g.\@}}
    \newcommand{\range}[2]{\ensuremath{#1-#2}}
    \newcommand{\unit}[1]{\ensuremath{\,\mathrm{#1}}}
    
    \newcommand{\micrometers}{\unit{\mu{}m}}
    \newcommand{\millimeters}{\unit{mm}}
    \newcommand{\centimeters}{\unit{cm}}
    \newcommand{\meters}{\unit{m}}
    \newcommand{\mwe}{\unit{m.w.e.}}
    \newcommand{\mbar}{\unit{mbar}}
    \newcommand{\Kelvin}{\unit{K}}
    \newcommand{\Megakelvin}{\unit{MK}}
    \newcommand{\mA}{\unit{mA}}
    \newcommand{\muA}{\unit{\mu{}A}}
    \newcommand{\emuA}{\unit{e\mu{}A}}
    \newcommand{\emA}{\unit{emA}}
    \newcommand{\pmA}{\unit{pmA}}
    \newcommand{\neV}{\unit{neV}}
    \newcommand{\eV}{\unit{eV}}
    \newcommand{\keV}{\unit{keV}}
    \newcommand{\MeV}{\unit{MeV}}
    \newcommand{\kV}{\unit{kV}}
    \newcommand{\MV}{\unit{MV}}
    \newcommand{\GHz}{\unit{GHz}}
    
    \newcommand{\minutes}{\unit{min}}
    \newcommand{\years}{\unit{y}}
    \newcommand{\Gigayears}{\unit{Gy}}

    \newcommand{\an}{\ensuremath{(\alpha,\mathrm{n})}}
    \newcommand{\ag}{\ensuremath{(\alpha,\gamma)}}
    \newcommand{\pg}{\ensuremath{(\mathrm{p},\gamma)}}
    \newcommand{\pa}{\ensuremath{(\mathrm{p},\alpha)}}
\begin{document}

\markboth{Aliotta et al.}{Exploring Stars in Underground Laboratories}

\title{Exploring Stars in Underground Laboratories: Challenges and Solutions}

\author{Marialuisa Aliotta,$^1$ Axel Boeltzig,$^2$ Rosanna Depalo,$^3$ and Gy\"orgy Gy\"urky,$^4$
\affil{$^1$SUPA, School of Physics and Astronomy, University of Edinburgh, Edinburgh, United Kingdom, EH9 3FD; email: m.aliotta@ed.ac.uk}
\affil{$^2$Department of Physics, Universit\`a degli Studi di Napoli ``Federico II'', Naples, Italy, 80126}
\affil{$^3$Universit\`a degli Studi di Milano and INFN Milano, Milano, Italy, 20133}
\affil{$^4$Institute for Nuclear Research (Atomki), Debrecen, Hungary, 4001, PO Box 51}}

\begin{abstract}
For millennia, mankind has been fascinated by the marvel of the starred night sky. 
Yet, a proper scientific understanding of how stars form, shine and die, is a relatively recent achievement, made possible by the interplay of different disciplines as well as by significant technological, theoretical and observational progress. 
We now know that stars are sustained by nuclear fusion reactions and are the furnaces where all chemical elements continue to be forged out of primordial hydrogen and helium.
Studying these reactions in terrestrial laboratories presents severe challenges and often requires developing ingenious  instrumentation and detection techniques.
Here, we reveal how some of the major breakthroughs in our quest to unveil the inner workings of stars have come from the most unexpected of places: deep underground. 
As we celebrate 30 years of activity at the first underground laboratory for nuclear astrophysics (LUNA), we review some of the key milestones and anticipate future opportunities for further advances both at LUNA and at other underground laboratories worldwide.
\end{abstract}



\maketitle

\tableofcontents


\section{INTRODUCTION}
\label{sec:intro}

Much of what we know about the makeup of the Universe stems from the study of the electromagnetic radiation emitted by stars, galaxies, the interstellar medium, and the early Universe some 380,000\,years after the Big Bang. 
Stellar spectra, in particular, provide a wealth of information about the physical properties of a star's outermost layers: like cosmic bar codes, their dark absorption lines\footnote{Absorption lines are caused by the absorption of the radiation emitted by a star at specific frequencies characteristic of the chemical elements present in the stellar atmosphere.} reveal the chemical composition of the gas clouds out of which a star formed.

Over the last decades, ground-based and space-borne telescopes have opened up unprecedented opportunities to observe the Universe at virtually all wavelengths, from radio-frequencies, to X- and $\gamma$-ray energies.
More recently, the direct detection of gravitational waves (disturbances in the curvature of space-time) resulting from colliding neutron stars and/or black holes has marked the birth of a multi-messenger era in astronomical observations \cite{Abbott2016,Bartos2017}.
Other direct messengers come in the form of tiny amounts of matter that reach the Earth from outer space, such as cosmic rays (highly energetic charged particles), pre-solar grains found in meteorite samples, and neutrinos (both from the Sun and nearby supernovae).
Like tiles of a cosmic jigsaw puzzle, all these pieces of information reveal a tremendous diversity of cosmic objects, but also point to a striking feature of the Universe as a whole:  98\% of all visible matter consists of hydrogen and helium, with all other chemical elements, from carbon to uranium (collectively called ``metals’’), making up a mere 2\%!

And yet, where do these elements come from? How, when and where were they forged? 
In an attempt to answer these questions, in 1957 Burbidge, Burbidge, Fowler and Hoyle (B$^2$FH) \cite{B2FH}, and independently Cameron \cite{Cameron1957}, laid the foundations of nuclear astrophysics, a vibrant interdisciplinary field that brings together astronomy, nuclear-, atomic-, and plasma physics, and requires efforts in observations, theory, and experiments.
The framework suggested to explain the origin of the chemical elements and their abundance distribution in the Universe has stood the test of time, through a wealth of astronomical observations and experimental verification.

Starting from protons and neutrons, the key building blocks produced during the Big Bang, all other elements are created via complex patterns of nuclear reactions. 
Some occurred within the first few minutes of life of the Universe and led to the synthesis of $^2$H (deuterium), $^3$He, $^4$He, as well as traces of Li, Be and B.
As the Universe continued to expand, nucleosynthesis halted for a few hundred million years, until the first stars (and galaxies of stars) started to emerge from the gravitational contraction of vast molecular clouds.
It is in the hot interiors of stars that heavier elements could finally be produced (and are still produced today) through sequences of fusion reactions between charged particles, starting from H and He.
As the binding energy per nucleon increases as a function of mass number up to a maximum in the iron mass region ($A\sim 60$), nuclear fusion liberates energy and stabilizes a star against further contraction. 
However, fusion beyond Fe becomes energetically disfavoured and different reaction mechanisms must be invoked to explain the origin of elements heavier than iron. 
These mechanisms involve a sequence of slow- or rapid- neutron capture processes (s- and r-process, respectively) followed by $\beta^-$ decays, where an excess neutron in a nucleus is converted into a proton, thus leading to a new chemical element.

Like gigantic cauldrons in the cosmos, stars are therefore responsible for ``cooking'' all elements out of primordial ingredients, H and He. 
Eventually, the more massive of stars end their lives in catastrophic supernova explosions, thus ejecting back into the interstellar medium all the new elements created in their interiors. 
In turn, new generations of stars, born out of this enriched cosmic soup, will too contribute to the chemical evolution of the Universe, in a continuous re-cycling of stellar material.

From these general considerations, it emerges that nuclear physics and nuclear properties take centre stage in the description of the chemical evolution of stars, galaxies and ultimately the Universe. 
A key goal of nuclear astrophysics consists in replicating in the laboratory the nuclear reactions that take place in stars, so as to better understand the intricate pattern of processes that have led to the distribution of abundances that we observe today.
Unfortunately, because of the relatively small energies at which nuclear reactions take place in stars, measuring their cross sections, \ie{} the probability for a reaction to occur, poses formidable challenges to experimenters. 

In this review, we briefly recall the main features of thermonuclear reactions in stars to illustrate the challenges and requirements related to their study in terrestrial laboratories. 
We then show how performing experiments in  underground laboratories has proved essential not just when studying reactions that liberate gamma rays, but also those that emit neutrons or charged particles.
Following the pioneering work performed at the first ever Laboratory for Underground Nuclear Astrophysics (LUNA), we celebrate its legacy 30\,years on, as we look forward to future opportunities, both at LUNA and elsewhere, to unveil some of the best kept secrets of nature.


\section{NUCLEAR REACTIONS IN STARS: 
Principles of stellar evolution and nucleosynthesis }
\label{sec:thermoreac}

Broadly speaking, quiescent (\ie{}, non explosive) stellar evolution proceeds through a sequence of gravitational contractions and distinct stages of nuclear fusion, during which new chemical elements are produced. 
Stars form from molecular gas clouds composed of $^1$H, $^4$He (75\% and 25\% by mass, respectively) and traces of heavier elements. 
As the gas cloud collapses, about half of its gravitational energy is radiated away while the other half is converted into heat, thus increasing the gas temperature. 
Provided the mass of the proto-stellar cloud is large enough, sufficiently high temperatures ($T \sim \range{10^6}{10^7}\Kelvin{}$) can be reached in the star's innermost regions to trigger nuclear fusion reactions. 
These liberate energy and stabilize the star against further gravitational contraction.
What happens next critically depends on the initial mass of the collapsing cloud and -- to a lesser extent -- its initial chemical composition.

The first, and longest, of the nuclear burning epochs, \emph{hydrogen burning}, converts hydrogen into helium over typical time scales of $10^{10}\years{}$, for a star like our Sun, to only $10^6\years{}$, for a star of about 40~solar masses ($M_\odot$).
After all hydrogen in the core has been exhausted, the star can no longer support the weight of its outer layers and gravitational contraction sets in once again further heating the stellar core.
Low-mass stars like our Sun will experience \emph{helium burning}, producing carbon and oxygen through the fusion of $^4$He nuclei (alpha particles) via the so-called $3\,\alpha$ process ($3\,\alpha \rightarrow {}^{12}\mathrm{C} + \gamma$) and the $^{12}\mathrm{C}\ag{}^{16}\mathrm{O}$ reactions. 
Low-mass stars will eventually die as white dwarfs, \ie{} highly dense and compact objects supported by electron degeneracy \cite{Rolfs1988}.
By contrast, more massive stars ($M > 8 M_\odot$) will evolve through more advanced stages of nuclear fusion ({\em carbon-, neon-, oxygen-, silicon-burning}), each leading to the synthesis of increasingly heavier elements up to Fe, before finally exploding as core-collapse supernovae.

In a stellar plasma, nuclear reactions are initiated by the thermal motion of nuclei and are therefore referred to as \emph{thermonuclear reactions}.
For a non-relativistic and non-degenerate plasma in thermodynamic equilibrium at a temperature $T$, the relative velocity distribution of nuclei is well described by the Maxwell-Boltzmann distribution, $\phi(v)dv = \phi(E)dE \propto \sqrt{E}\exp(-E/kT)$, where $k=8.6173\times10^{-5}$~eV/K is the Boltzmann constant.
However, the energy $kT$, at which the velocity distribution reaches its maximum, turns out to be always smaller than the repulsive Coulomb barrier between the interacting nuclei. 
For example, at temperatures $T\simeq 15\times 10^6 \Kelvin$ (as in the centre of the Sun) the average energy of the ions in the plasma is $kT \simeq 1.3\keV$, \ie{} much lower than the Coulomb repulsion between even the lightest of charges, namely two protons ($E_{\rm Coul}=0.5\MeV$).
It follows that nuclear reactions in stars are only possible thanks to the quantum-mechanical tunnelling through the barrier.
In the absence of a centrifugal barrier that would further hinder a fusion reaction, the tunnelling probability can be described by an exponentially decreasing function of energy as $\exp({-2\pi\eta})$, with $2\pi\eta = 31.29 \ Z_1 Z_2 (\mu/E)^{1/2}$ ($Z_i$ being the atomic numbers of the interacting particles, $\mu$ their reduced mass in atomic mass units, and $E$ the centre-of-mass energy in kiloelectronvolt (keV) units \cite{Rolfs1988}).

The cross section $\sigma(E)$ for a {\em non resonant} reaction can then be expressed as:
\begin{equation}
	\sigma(E) = \frac{1}{E}\exp(-2\pi\eta)\,S(E)
    \label{eq:cross-section}
\end{equation}
where, $1/E \propto \pi \lambdabar^2$ is a non-nuclear term involving the de~Broglie wavelength of the interacting nuclei, and the {\em astrophysical $S(E)$ factor}, defined by this equation, contains all the strictly nuclear physics effects of the interaction.
Note that for non-resonant reactions, the astrophysical $S(E)$ factor varies little with energy. 
\begin{marginnote}
	\entry{Non resonant reactions}{Also known as {\em direct reactions}, they occur without the formation of an intermediate {\em compound} nucleus and proceed directly into a final nucleus with the prompt emission of either a $\gamma$ ray or a particle. }
\end{marginnote}
\begin{marginnote}
	\entry{Resonant reactions}{Can be regarded as a two-step process and often proceed through an excited state of the compound nucleus. Their cross sections show pronounced peaks and can be expressed by the Breit-Wigner formula (see for example \cite{Rolfs1988}.)}
\end{marginnote}

\begin{figure}[htb]
	\includegraphics[width=\textwidth]{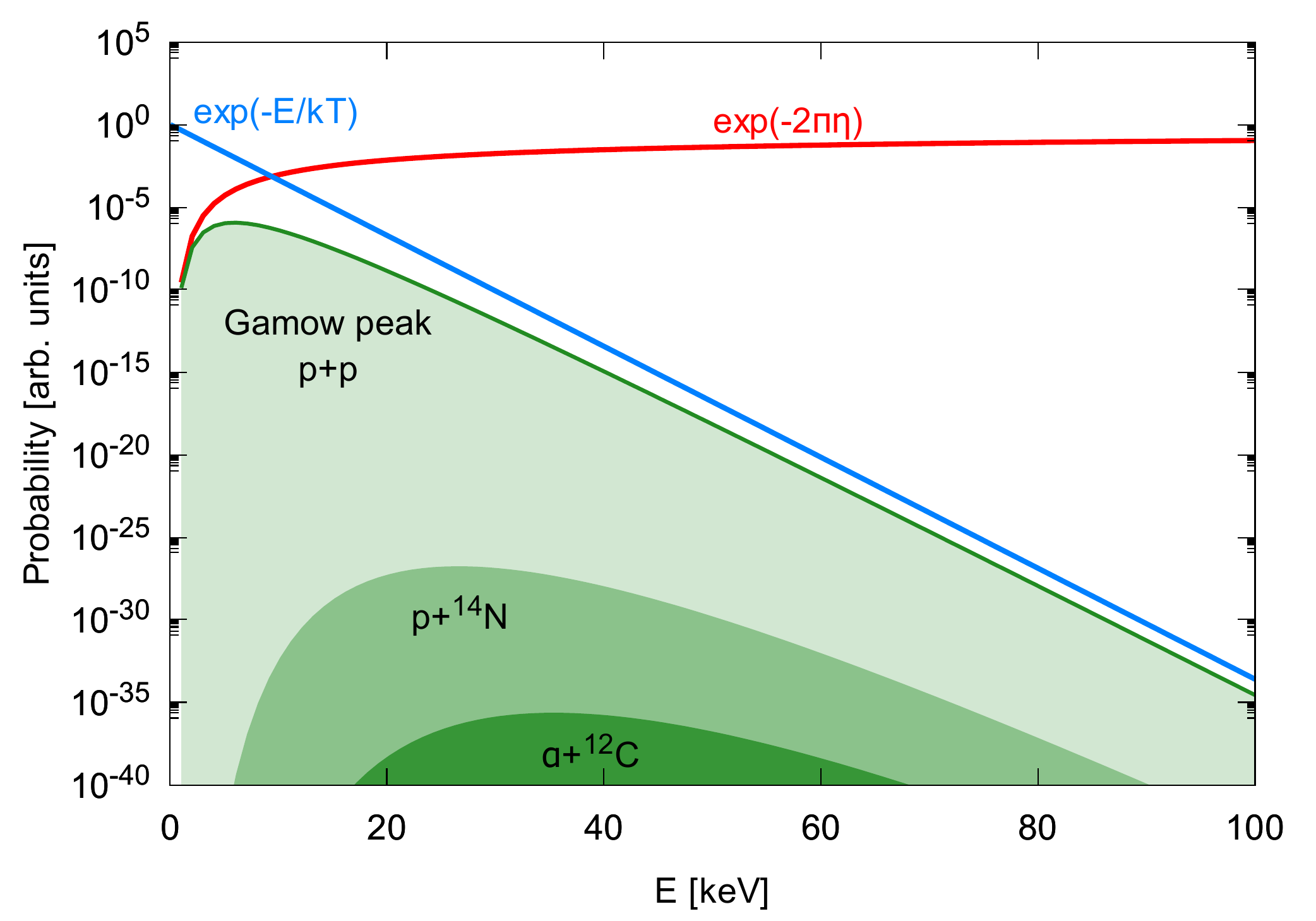}
  	\caption{(Color online) 
  	The Gamow peak curve (green) arises from the product of the Maxwell-Boltzmann distribution at a temperature $T$ (blue) (here $T = 15\times10^6 \unit{K}$) and the tunnelling probability between the interacting nuclei (red). 
  	The integral of the peak (shaded area) is proportional to the reaction rate.
  	Note how, at the same temperature, the Gamow peak curves shift to higher energies for reactions between heavier charges and become progressively smaller because of the correspondingly lower tunnelling probabilities $\exp(-2\pi\eta)$ (shown here only for the p+p case).}
    \label{fig:gamow-peak}
\end{figure}

The key quantity of interest for astrophysical purposes is not just the cross section, but the {\em reaction rate} per particle pair, defined as:
\begin{equation}
	\left<\sigma v\right> = \int_0^\infty \phi(v)\,v\,\sigma(v)\,\mathrm{d}v = \int_0^\infty \phi(E)\,v\,\sigma(E)\,{\rm d}E
	\propto \int_0^\infty S(E) \exp\left[-E/kT-2\pi\eta \right]{\rm d}E
	\label{eq:rate}
\end{equation}
with $v$ being the relative velocity between the interacting particles.  
The product of the Maxwell-Boltzmann distribution and the tunnelling probability gives rise to a peak-shaped curve, commonly referred to  as the \emph{Gamow peak} (as shown in {\bf Figure \ref{fig:gamow-peak}}), which represents the energy region at which a given reaction is most likely to occur or, equivalently, the energy region that maximises the reaction rate per particle pair.  
To a first approximation, the reaction rate $\left<\sigma v\right>$ is proportional to the area under the Gamow peak (see \cite{Rolfs1988} for a more detailed and rigorous account of this derivation) and, at any given temperature, its value drops dramatically by several orders of magnitude as the charges of the interacting particles increase (\textbf{Table~\ref{tab:coulomb}}). 
Incidentally, this explains why stellar evolution proceeds through well-defined stages of nuclear burning: 
nuclear reactions involving heavier and heavier elements  require ever increasing temperatures, which can only be attained through the gravitational contraction that follows the exhaustion of a previous nuclear fuel in the core of the star.

As the star evolves, its temperature changes and so will the location and width of the Gamow peak of a given reaction, as well as the associated astrophysical reaction rate $\left<\sigma v\right>$ ({\bf Figure \ref{fig:gamow-peak}}). 
Stellar models and nucleosynthesis network calculations therefore require the accurate knowledge of hundreds of nuclear reaction cross sections, each over a broad energy region.
Unfortunately, because of our limited knowledge of nuclear forces and interactions it is impossible to accurately predict from first principles nuclear reaction cross sections as a function of energy. 
Instead, these must be determined experimentally.
In the next section we explore the main challenges and the key requirements for laboratory measurements of reactions between stable nuclei at the energies of astrophysical interest.

\begin{table}[t]
\begin{center}
	\caption{Coulomb barriers and Gamow peak energies for nuclear reactions between nuclei with increasing charge numbers, at a temperature $T = 15 \times 10^6\Kelvin{}$. The last column represents the area under the Gamow peak, a proxy for the corresponding nuclear reaction rates. \\}
	\label{tab:coulomb}
	\begin{tabular}{llll}
	\hline 
	reaction & Coulomb barrier &  Gamow peak  & peak integral $\propto \left< \sigma v \right>$ \cite{Rolfs1988}\\
	\hline
	p + p 	& 0.55 MeV	& 5.9 keV 	& $1.1 \times 10^{-6}$ \\
	p + $^{14}$N 	& 2.27 MeV	& 6.8 keV 	& 
	$1.8 \times 10^{-27}$ \\
	$\alpha$ + $^{12}$C & 3.43 MeV 	& 56 keV 	& $3.0 \times 10^{-57}$ \\
	$^{16}$O + $^{16}$O & 14.07 MeV	& 237 keV 	& $6.2 \times 10^{-239}$ \\
	\hline
	\end{tabular}
\end{center}
\end{table}


\section{NUCLEAR REACTIONS IN THE LABORATORY: Challenges and requirements}
\label{sec:lab}

Nuclear reactions of astrophysical interest can be studied in the laboratory using a particle accelerator capable of delivering a beam of ions (often protons or $\alpha$-particles) onto a suitable gas or solid-state target over a broad energy range. 
The target is surrounded by detectors sensitive to the type of radiation (\ie{} $\gamma$ rays, charged particles or neutrons) produced by the reaction under study.
Ideally, reaction cross sections should be measured directly at the Gamow peak for the given pair of interacting nuclei and at the relevant stellar temperature. 
Sadly, this is easier said than done.
Because of the steep exponential drop associated with the tunnelling probability (eq.~\ref{eq:cross-section}), cross sections are typically in the range of pico- to femto-barn ($1\unit{b} = 10^{-24}\centimeters^2$) and translate into extremely low counting rates, ranging from a few counts per hour to a few counts per year in the most extreme cases!
Therefore, cross-section evaluations at stellar temperatures often rely on extrapolations from data taken at higher energies. 
While such extrapolations are often performed with the aid of theoretical formalism, such as the $R$-matrix \cite{Descouvemont2010}, uncertainties often remain because of possible contributions from unknown resonances (either above or below thresholds) and because of the unclear influence of the electron screening effect \cite{Assenbaum1987} at the lowest interaction energies.

To improve on cross-section extrapolations, considerable effort must be devoted to pushing direct measurements to lower and lower energies, which however poses considerable challenges and requires a number of expedients to improve the signal-to-noise ratio of the reaction under study. 
This is generally achieved by a combination of measures aimed at improving the signal (\eg{} through maximising beam currents, target density, and detection efficiencies), and reducing the noise (\ie{} the background signal) both natural and beam-induced.
These measures are discussed in the following sections, together with a few additional requirements not normally encountered in ordinary nuclear physics experiments.
\begin{marginnote}
\entry{Electron screening}{At extremely low energies, the presence of electrons around the positively charged nuclei (both in stellar plasma and in the laboratory) effectively reduces the Coulomb repulsion and enhances the cross section. 
Measured cross sections must be corrected for the screening effect before they can be used for astrophysical purposes.
Unfortunately, tension remains between theoretical predictions and laboratory measurements of the electron screening (e.g. \cite{Raiola2004}), leading to uncertain corrections.}
\end{marginnote}

\subsubsection*{Long term stability of accelerators and targets.}
The low cross sections of astrophysical reactions necessitate high beam intensities and long irradiation times of up to several months. 
The long term stability of both accelerators and targets is thus crucial. 
Modern accelerators provide good energy stability over time but dedicated tests are often required for nuclear astrophysics applications. 
Highly automatized systems allowing for unattended, remote operation are also useful, but require reliable safety solutions. 

Long irradiation times at high beam currents may easily result in strong target degradation in the case of solid state targets. 
Sputtering of the target material, heat-induced effects, as well as implantation of the beam particles may alter the target properties. 
Thus, the choice of the target material and the preparation method must be optimized for target stability and target degradation must be kept under control. 
Before the actual cross section measurements, fresh targets are usually characterized through suitable ion beam analysis methods \cite{Gyurky2019b} to accurately determine their thickness, composition and stoichiometry (in case of compounds). 
Repeated measurements of these target properties in the course of the experiment are used to monitor the target degradation and keep it at an acceptable level \cite{Ciani2020}. 
Depending on the precision required for the cross section measurement, some degree of degradation -- if precisely known -- may be tolerated, but the replacement of spent targets with fresh ones is eventually needed after a certain amount of charge has been accumulated. 

\subsubsection*{Accurate knowledge of interaction energy.}

In addition to the energy stability (in time) of the ion beam, a precise knowledge of its absolute energy and energy spread is also necessary at sub-Coulomb energies because the steep energy dependence of non-resonant cross sections sensitively affect the stellar reaction rate $\left<\sigma v\right>$.
For example, in the $^3$He\ag{}$^7$Be reaction,
a $1 \keV$ beam energy uncertainty at $E_{\alpha} =220 \keV$ (a $0.5\%$ error) translates into a $4\%$ uncertainty in the cross section.
Similarly, the rate of reactions dominated by narrow resonances depends exponentially on the energy of the resonant state \cite{Rolfs1988}.
Thus, a precise energy calibration of the accelerator beam energy must be carried out.
This is typically achieved by exploiting suitable nuclear reactions which exhibit resonances of very well known energy. 
Depending on the dynamic range of the accelerator, different reactions may be exploited; for example, the $^{23}$Na\pg{}$^{24}$Mg, $^{25}$Mg\pg{}$^{26}$Al and $^{26}$Mg\pg{}$^{27}$Al reactions have all been used below $400\keV$ to a precision of $\pm 300\eV$ \cite{Formicola2003}.
Higher energy accelerators can make use of neutron threshold reactions \cite{Rajta2018} or other well-known resonances, \eg{} in the $^{27}$Al\pg{}$^{28}$Si reaction. 

The accurate knowledge of interaction energy not only poses requirements on the accelerator itself, but also on the target properties. 
As the beam loses energy when traversing the target, the cross section is effectively measured not at a single energy, but over a finite energy range. 
Information about the target thickness and composition is thus needed in order to calculate the effective energy of the measured cross section. 
This holds for both solid state and gas targets. 
For gas targets the so-called beam heating effect (\ie{} the increase in gas temperature due to the power deposited by the beam) further complicates the energy loss determination and needs to be carefully studied \cite{Marta2006}. 
Extended gas targets often have physical dimensions larger than the size of typical detectors. 
Since the detection efficiency may depend strongly on the position at which the nuclear reaction  occurs, the energy loss of the beam in the target gas must be convoluted with the detector efficiency in order to determine the effective interaction energy and its associated cross section  \cite{Bemmerer2018}.

\subsubsection*{Detectors.}

Requirements for a detection setup include a high efficiency and sensitivity to the radiation produced by the reaction under study, and -- at the same time -- a high signal-to-background ratio. 
The latter requirement can be achieved either by reducing backgrounds seen in the detector (as discussed below), or by optimising the detected signal, \eg{} through  improved energy resolution, pulse-shape discrimination, or the use of detectors with independent sensitive volumes to allow for powerful coincidence or anti-coincidence configurations.
Once again, the long data-taking times of low-yield experiments call for a long-term stability of the detection apparatus and related electronics, so frequent calibration measurements are needed to periodically check on the detector response. 
This is particularly true for low-background experiments where low statistics prevents identification of the region of interest from the measurement itself. 

\subsubsection*{Backgrounds.}

So far, we have looked at ways to maximise the counting rate by improving the ``signal'' through increased beam intensity and detection efficiency. 
However, the most insidious limitation to nuclear astrophysics experiments often arises from background signals that can mimic the signature of the reaction of interest but are caused in fact by different processes and therefore limit the sensitivity of the experiment.
Techniques to mitigate the various sources of background are highly specific to the type and origin of the background and typically include material screening and surface cleaning, active or passive shielding, and the rejection of background signals by pulse-shape discrimination techniques \cite{Cuesta2013}.

Background sources can be grouped according to their origin as beam-induced, intrinsic, and environmental.
In accelerator-based experiments, beam-induced backgrounds originate from spurious interactions of the ion beam with the experimental setup. 
These may occur in the target material, or at any other point along the path of the beam.
Reactions on trace contaminants in the target material (often light elements with their low Coulomb barrier) can be reduced by choosing chemically pure materials for target production, or by treating these materials through etching to reduce surface contaminants \cite{Caciolli2012,Champagne2014}, or through heating to drive out contaminants from the bulk of the target \cite{Depalo2021}. 
Similar considerations apply to other components along the beam path (\eg, apertures or collimators) which should typically be chosen to have a large atomic number (\eg, tungsten) and be frequently cleaned. 
For solid targets, careful handling and storage (\eg, under protective atmosphere to avoid oxidation or absorption of humidity) may reduce the risk of target contamination in the time between target production and measurement.

Intrinsic backgrounds are present in the experimental setup, most critically in the detection material itself. 
Radioactive nuclides in the detector medium may cause background signals through the emission of radiation in their decay. 
These radioactive nuclides may be part of a decay chain originating from long-lived primordial nuclides. 
Nuclides with shorter life times may be the result of artificial radioactivity (\ie{} produced in man-made processes), or cosmogenic activation (\ie{} produced through interactions with cosmic rays).
For geometrical reasons, the closer this radiation is to the sensitive volume of the detector, the more likely it will result in a background signal. 
This is especially true for charged particle radiation because of its short range in matter. 

Finally, environmental (or ambient) backgrounds are caused by the radiation field at the location of the experiment, which is present independently of the experimental setup. 
This includes natural radioactivity, as well as the effects produced by cosmic radiation. 
These sources of environmental radiation may result in background signals through their direct interaction with the detector setup. 
For example, characteristic $\gamma$-ray energies from the decay of radionuclides in the long-lived decay chains are readily observed in a high-resolution $\gamma$-ray detector. 
Environmental radiation may also contribute to the detector background through secondary effects, such as \an{} reactions induced by the $\alpha$ particles produced in a radioactive decay and leading to secondary neutrons. 
Environmental $\gamma$-ray background is often reduced by shielding the detection setup with active and/or passive high-purity and high-Z materials (typically lead and copper).
Yet, the best and most effective way of reducing a major source of natural background consists in performing experiments deep underground, where the influence of cosmic rays can be greatly reduced. 
The validity of this approach was dramatically demonstrated with the installation of a small accelerator specifically designed for nuclear astrophysics studies.

In the following section, we describe how the many requirements presented here have been successfully met at the Laboratory for Underground Nuclear Astrophysics (LUNA), the first laboratory of its kind worldwide. 
After summarising its main components and instrumentation, we will show how the underground location has proved instrumental not just for studying reactions producing $\gamma$ rays, but also for those emitting charged particles or neutrons.


\section{NUCLEAR ASTROPHYSICS UNDERGROUND: 
The LUNA facility}
\label{sec:luna}

\subsubsection*{Accelerators.}
The Laboratory for Underground Nuclear Astrophysics (LUNA) was established in 1991 with the installation of a $50\kV$ accelerator at the Laboratori Nazionali del Gran Sasso (LNGS) of the Italian Institute of Nuclear Physics (INFN). 
The accelerator was specifically designed to investigate nuclear reactions from the pp chain at energies close to the solar Gamow window\footnote{By approximating the Gamow peak curve to a Gaussian function, the Gamow window can be defined as the 1/$e$ width $\Delta$ of the Gaussian function. The Gamow window therefore represents the energy region that makes the most contribution to the area under the Gamow peak (see \cite{Iliadis2007} for a rigorous derivation).} \cite{Greife1994}. 
Thanks to the success of those early measurements (see {\bf Section \ref{sec:luna-overview}}), the pilot accelerator was replaced in 2001 with a $400\kV$ electrostatic machine, which is still in operation today \cite{Formicola2003}. 
Its dynamic range has allowed for the investigation of many key reactions at the relevant energies of hydrogen burning in different phases of stellar evolution, including during the main sequence, the Red Giant Branch (RGB) and Asymptotic Giant Branch (AGB) stages, classical novae explosions and Big Bang Nucleosynthesis (BBN) (see {\bf Section \ref{sec:luna-overview}} for key scientific highlights).
    
In the LUNA-$400\kV$ accelerator the high voltage is produced by an inline-Cockcroft-Walton power supply capable of handling currents as high as $1\mA$ at $400\kV$ \cite{Formicola2003}. 
A radio-frequency source provides $1\mA$ H$^{+}$ beams and $500\muA$ He$^{+}$ beams. 
Once extracted, the beam is accelerated and redirected by 45$^\circ$ magnets, either towards the gas or towards the solid target station. 
The absolute beam energy is calibrated to a precision of $\pm 300\eV$, the proton energy spread is lower than $100\eV$ and the long-term energy stability is $5\eV\,\mathrm{h}^{-1}$ \cite{Formicola2003}. 
In addition, a number of safety interlocks allows the machine to operate in stable conditions even without human supervision. 
These characteristics make LUNA-$400\kV$ the ideal machine for nuclear astrophysics experiments requiring long data taking periods.

\subsubsection*{Beam lines: solid- and gas-target stations.}

\begin{figure}[t]
	\includegraphics[width=\textwidth]{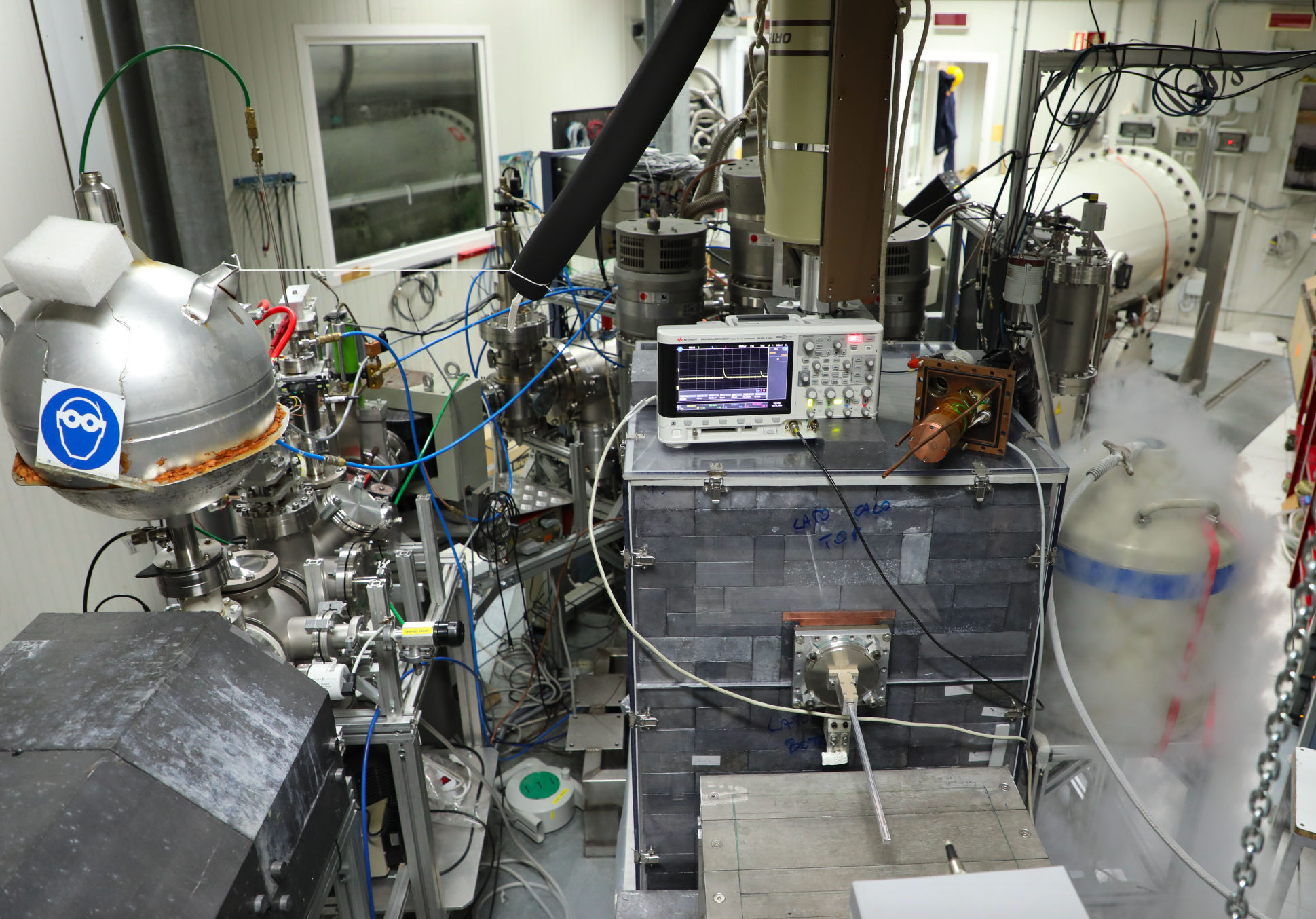}
    \caption{(Color online) Picture of the LUNA-400kV accelerator (in the background, top right) and the two beam lines (in the foreground). Both solid-target (left) and gas-target (centre) setups are surrounded by massive lead shields for further background suppression.}
    \label{fig:LUNA-setup}
\end{figure}

The LUNA-$400\kV$ accelerator is equipped with two beam lines: one hosting a windowless gas target system, the other hosting a solid target station, as shown in {\bf Figure  \ref{fig:LUNA-setup}}.
Thanks to the two setups it is possible to work with a wide range of target materials. 
Also, some nuclear reactions can be studied independently both with the gas target and with solid targets, checking for possible systematic effects due to target properties. 
As mentioned in {\bf Section~\ref{sec:lab}}, gas targets have the advantage of being stable upon irradiation with intense ion beams, but  require complex pumping systems, more elaborate techniques to determine the beam current, and precise density profile  measurements.  
Instead, solid targets are comparatively easier to handle, but they need to be periodically checked for degradation due to beam irradiation.
    
The LUNA windowless gas target system consists of three differential pumping stages. 
The absence of windows to confine the gas preserves the beam energy distribution before entering the target chamber. 
A continuous gas flux adjusted by a feedback system allows to keep a constant pressure in the range of $\range{0.1}{10}\mbar$ inside the target chamber. 
The gas is continuously pumped from the chamber through three pumping stages that gradually bring the pressure down to the $10^{-7}\mbar$ range. 
The pumping stages are separated by long water cooled apertures of decreasing diameter serving both to collimate the beam and to increase the impedance for the gas flow from the target to the pumping stages.
The gas taken from the pumping stages can be either discharged or collected, purified and recirculated back into the gas target. 
Inside the target chamber, the beam is stopped on a beam calorimeter providing continuous and accurate beam current measurements. 
The target chamber geometries can be adapted to couple the gas target with different types of detectors.

The second beam line ends with a solid target station. Different vacuum chambers have been designed over the years to meet the requirements of individual experiments. 
In general, the target backings used are characterized by high purity levels and are thick enough to stop the beam. 
The target backing is therefore directly water-cooled to dissipate the power deposited by the beam. To reduce contaminant deposition on target, the vacuum chamber is equipped with a copper pipe cooled to liquid nitrogen temperature. 
The pipe is electrically insulated from the chamber and biased to typical voltages of $-300\unit{V}$ to suppress secondary electrons produced when the beam hits the target. 
Chamber and target holder are electrically insulated from the beam line, and in this configuration act as a Faraday cup for beam current integration.
    
\subsubsection*{Detectors.}

Various detectors have been used at LUNA, each with a dedicated target chamber. 
For the detection of $\gamma$ rays, high-resolution spectroscopy is performed with large-volume high-purity germanium detectors (HPGe). 
All LUNA HPGe detectors are made from materials with low intrinsic background so as to preserve the advantages of being underground. 
Over the years, different target-detector geometries have been adopted, depending on the requirements of the reaction under study. 
In addition, when $\gamma$ rays with energies lower than 3\MeV{} need to be detected, thick passive shielding made of lead and copper are used to suppress the environmental background due to naturally-occurring radioactive isotopes (see also {\bf Section \ref{sec:background}}).
If extremely weak cross sections are to be measured and the sensitivity needs to be pushed to the limit, a large-volume Bismuth Germanate (BGO) detector covering almost the full solid angle around the target is used. 
The detector is optically segmented in six sections, with each crystal coupled to a photomultiplier. 
The BGO detector has low energy resolution, but can be used as a calorimeter by summing the energies of all coincident $\gamma$ rays detected in any of its sections. 
With this approach, only one peak is observed in the spectrum at the nuclear excitation energy, and the detection efficiency is of the order of 60\%. 
Lately, the BGO detector has also been used successfully to determine $\gamma$-decay branching ratios exploiting $\gamma$-$\gamma$ coincidences in sections pairs.

More recently, the LUNA collaboration has expanded its detector suite with the addition of an array of large area silicon detectors for charged-particle detection ({\bf Section \ref{sec:bkg_part}}), 
and a set of $^3$He counters for neutron detection ({\bf Section \ref{sec:bkg_neu}}). 

In the following we present three examples of recent studies that demonstrate how going underground has been instrumental not just for the detection of $\gamma$ rays, but also for the detection of charged-particles, and neutrons.

\subsection{Background suppression deep underground}
\label{sec:background}

\subsubsection{Gamma-ray detection: The $^2$H\pg{}$^3$He reaction}
\label{sec:bkg_gammas}
    
\begin{figure}[t]
	\includegraphics[width=\textwidth]{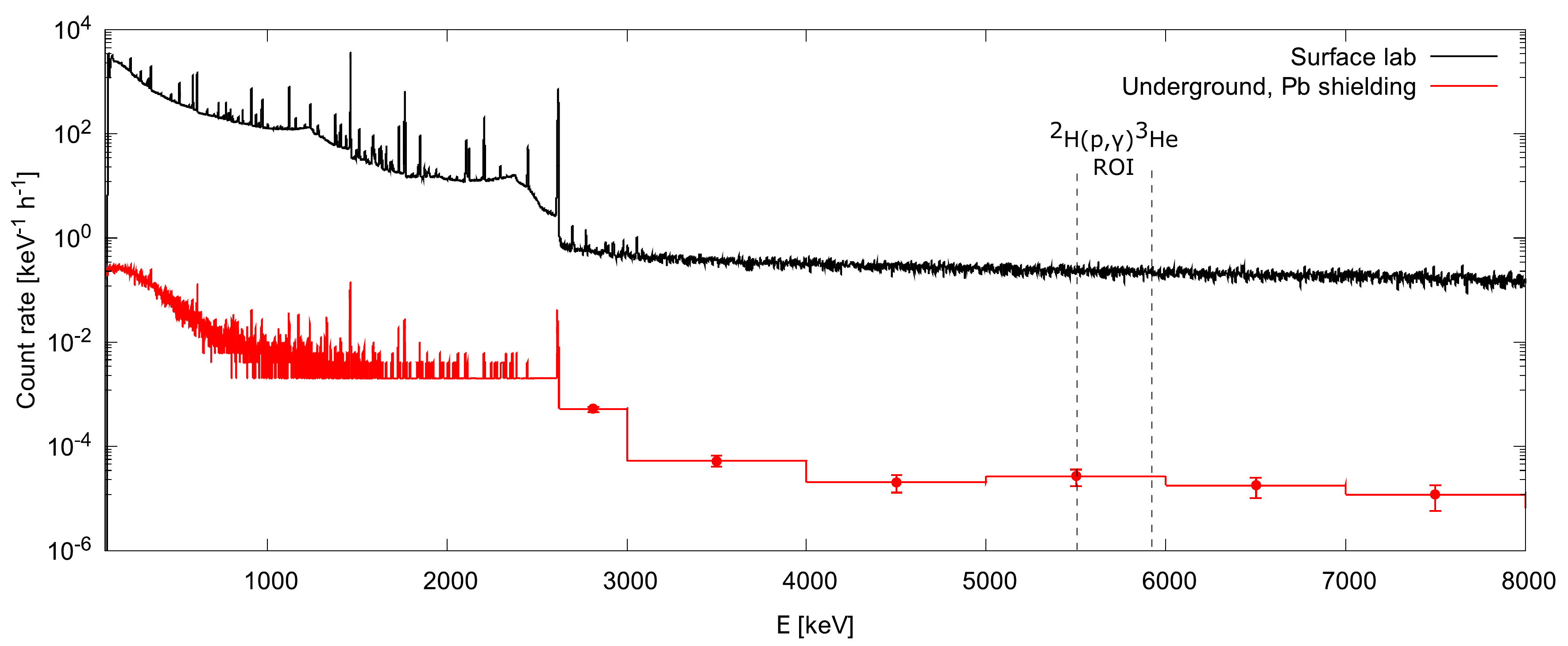}
    \includegraphics[width=\textwidth]{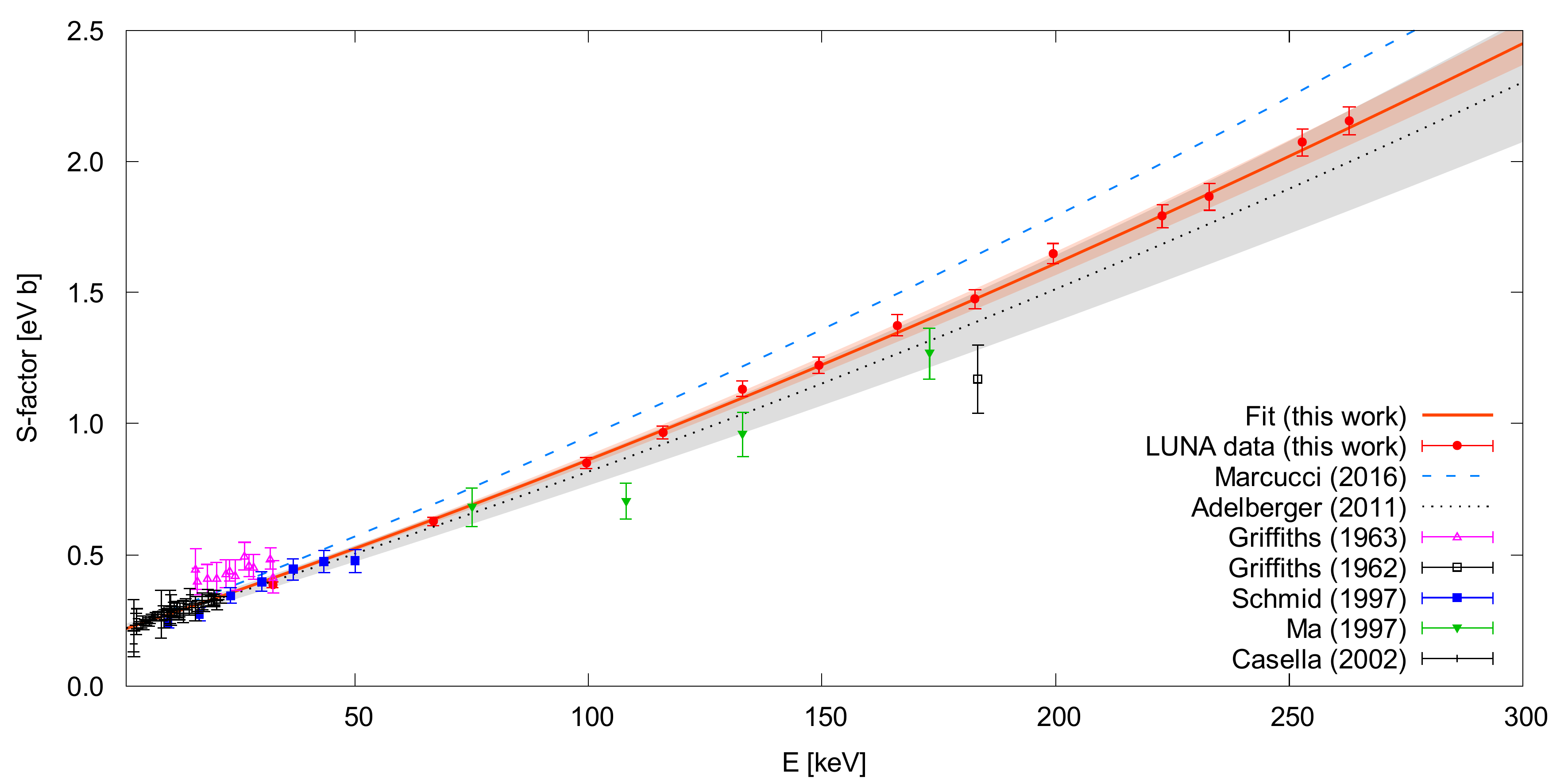}
	\caption{(Color online) Upper panel: Background spectra recorded with a HPGe detector on the Earth's surface (black) and underground at LUNA, with the copper and lead shielding described in \cite{Cavanna2014}. Lower panel: 
       Astrophysical S-factor of the $^2$H\pg{}$^3$He reaction. The LUNA results are compared to the literature.}
    \label{fig:Background_gamma}
\end{figure}

The study of nuclear reactions emitting $\gamma$ rays is especially favourable underground because of the natural background suppression afforded by the rock overburden.
The $1.4\unit{km}$ ($3800\unit{meters}$ of water equivalent, m.w.e.) of rocks above LNGS, for example, leads to a six-order-of-magnitude suppression of the cosmic-induced background at $\gamma$-ray energies above $3\MeV$, as shown in {\bf Figure \ref{fig:Background_gamma}} (upper panel). 

At lower energies, the $\gamma$-ray background is dominated by $\gamma$ rays emitted in the decay of naturally-occurring radioactive isotopes. 
However, also in this energy region a deep  underground location brings a substantial advantage. 
This component of the $\gamma$-ray background is typically suppressed by surrounding the detector with passive shielding made of high-purity and high-Z materials (typically lead and copper).  
On the Earth surface, the thickness of the shielding is limited by the fact that the interaction of cosmic rays within the shielding itself produces radioactive isotopes and secondary radiation. 
This problem is significantly reduced underground, where a much thicker shielding can then be used.

As an outstanding example of the potential of underground experiments aimed at detecting $\gamma$ rays we present here the case of the $^2$H\pg{}$^3$He reaction ($Q = 5.5\MeV$). 
This reaction plays a key role in the first minutes of the life of our Universe, as it contributes to deuterium destruction during Big Bang Nucleosynthesis (BBN). 
In BBN studies, deuterium abundance is used as an indicator of cosmological parameters, since it is particularly sensitive to the baryon density, or alternatively the baryon-to-photon ratio, of the Universe. 
Until recently, however, the $^2$H\pg{}$^3$He cross section represented the main source of uncertainty on the predicted abundance of primordial deuterium \cite{Planck2018}. 
The uncertainty was due to the lack of experimental data at the energies of interest for BBN ($\range{30}{300}\keV$), where only two data sets were available in the literature \cite{Ma1997,Tisma2019}, both with relatively high systematic errors. 
\begin{marginnote}[]
	\entry{Big Bang Nucleosynthesis}{BBN refers to a network of reactions taking place soon after the Big Bang and leading to the synthesis of light elements such as deuterium, $^3$He,  $^4$He, and $^7$Li.}
\end{marginnote}
The $^2$H\pg{}$^3$He cross section was also evaluated by \emph{ab-initio} calculations \cite{Marcucci2016}, but the difference between theoretical calculations and experimental data led to some ambiguity in the choice of the cross section for BBN models and resulted in a high uncertainty on the inferred cosmological parameters \cite{Planck2018}.

At LUNA, the $^2$H\pg{}$^3$He reaction was studied at centre-of-mass energies between 30 and $263\keV$ using the windowless gas target system and an HPGe detector mounted in close geometry \cite{Mossa2020-EPJA}. 
The reaction cross section was measured with an unprecedented low systematic error ($\leq 3$\%) \cite{Mossa2020} (see lower panel in {\bf Figure \ref{fig:Background_gamma}}). 
These results allowed us to significantly reduce the uncertainty on BBN predictions of the baryon density and the effective number of neutrino families \cite{Mossa2020}. 
Thanks to the new LUNA data, the cosmological parameters provided by BBN models are now in better agreement with those derived from the analysis of the CMB anisotropies, thus supporting the standard cosmological model \cite{Mossa2020}.

\subsubsection{Charged-particle detection: The $^{17,18}$O\pa{}$^{14,15}$N reactions}
\label{sec:bkg_part}

In a silicon semiconductor, the energy spectrum of cosmic muons shows a maximum near zero and decreases exponentially with energy \cite{Misiaszek2013}.
As cosmic muons are significantly suppressed underground, one can expect improved signal-to-background ratios also in deep-underground experiments aimed at detecting charged particles with silicon detectors.
Indeed, improved background suppression has been demonstrated in the study of the $^{17,18}$O\pa{}$^{14,15}$N reactions \cite{Bruno2016,Bruno2019}, which play an important role in the nucleosynthesis of key isotopes used to constrain stellar models of novae, AGB, and post-AGB stars.
Both reactions were studied at LUNA using an intense proton beam onto solid Ta$_2$O$_5$ targets enriched in either $^{17}$O or $^{18}$O. 
The main goals were to measure: 1) the strength of the $E_{\rm p} = 70\keV$ resonance in the $^{17}$O\pa{}$^{14}$N reaction, and 2) the excitation function of the $^{18}$O\pa{}$^{15}$N reaction at energies $E_{\rm p} = \range{60}{360}\keV$, to determine the strengths of four resonances of astrophysical interest.
Because of the low beam energies required, the kinematics of the emitted $\alpha$ particles was essentially governed by the  $Q$-values of the $^{17}$O\pa{}$^{14}$N and $^{18}$O\pa{}$^{15}$N reactions (1.192 and $3.98 \MeV$, respectively).

Thus, a purpose-built scattering chamber was developed to detect low-energy ($E_\alpha \simeq 200\keV$ and $2.3\MeV$, respectively) $\alpha$ particles with maximum efficiency.
The setup consisted of an array of six passivated implanted planar silicon detectors (thickness: $\range{300}{700} \micrometers$, active area: $9 \centimeters^2$) arranged over two rows at angles of $135.0^{\circ}$ and $102.5^{\circ}$, with an overall efficiency of 15\% \cite{Bruno2015}.
Each detector was protected by aluminized mylar foils of appropriate thickness, carefully chosen to suppress the large flux of elastically scattered protons, while allowing for the passage of the $\alpha$ particles with minimal energy loss. 
\begin{marginnote}
	\entry{AGB stars}{Over 90\% of stars with $M \geq 1~M_\odot$ will evolve into the Asymptotic Giant Branch phase, characte\-ri\-zed by a luminosity up to $10^4$ times and a factor of 2 lower surface temperature that those of the sun \cite{Vassiliadis1993,Marigo2008}.
	Their names arise from the position of the star in the Hertzsprung-Russell diagram.}
\end{marginnote}

\begin{figure}[t!]
	\vspace{-2.8cm}
    \includegraphics[angle=-90,width=\textwidth]{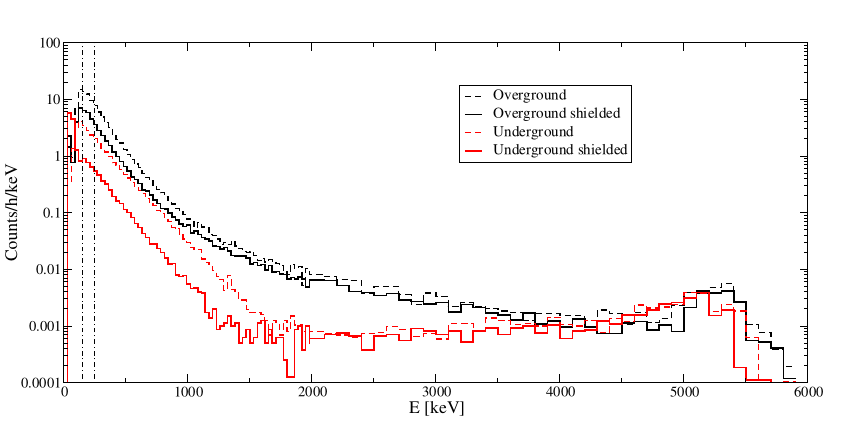}
	\vspace{-2.2cm} 
	\caption{(Color online) Background spectra obtained with passivated implanted silicon detectors overground (black lines) and underground (red lines) with and without lead shielding (solid and dashed, respectively) \cite{Bruno2015}. The vertical lines indicate the region of interest (ROI) for the detection of low energy $\alpha$ particles from the $^{17}$O\pg{}$^{14}$N reaction.}
    \label{fig:Bck_particles}
\end{figure}

In order to quantify the background reduction underground,  measurements were performed both overground (in Edinburgh, UK) and underground (at LUNA) using the same setup and electronics, both with and without a shield of lead bricks arranged around three sides of the chamber.
The results \cite{Bruno2015} are shown in {\bf Figure \ref{fig:Bck_particles}}. 
At energies up to about $1.5\MeV$, the spectra are dominated by a long exponential tail that is suppressed by a factor of about 15 for the underground shielded setup compared to the overground (un-shielded) measurement. 

The strongest background suppression (a factor of $\sim 23$) is observed around $2\MeV$, before gradually reducing to a factor of $\simeq 2$ at $4\MeV$. 
In the energy region $E \simeq \range{2}{4}\MeV$, the lead shielding does not appear to have a major effect either overground or underground. 
Although the source of background in this region is not obvious, we note that the contribution from  the low-energy tail of the broad peak a $E \sim 5.5\MeV$ is especially important underground.
This broad peak is likely due to the intrinsic activity of the silicon detectors since its contribution is not affected by either the lead shielding or the underground environment.

The combined effect of the  background suppression underground and the generally improved experimental conditions \cite{Bruno2015} have led to the most accurate value to date for the $E_{\rm p} = 70\keV$ resonance strength $\omega \gamma$ in $^{17}$O\pa{}$^{14}$N ($\omega \gamma = 10.0 \pm 1.4_{\rm stat} \pm 0.7_{\rm syst}\neV$). In turn, this has led to a factor-of-two improvement in the $^{17}$O\pa{}$^{14}$N reaction rate and to a reduced $^{17}$O/$^{16}$O ratio with important consequences for the origin of some oxygen-rich Group II pre-solar grains \cite{Lugaro2017}.
Similarly, we have obtained improved results on the $^{18}$O\pa{}$^{15}$N cross sections and resonance strengths, with tighter constraints on oxygen isotopic ratios \cite{Bruno2019}.
Future \pa{} measurements are foreseen at the LUNA-400kV accelerator in the next three-year scientific program (2022-24).

\subsubsection{Neutron detection: The $^{13}$C\an{}${}^{16}$O reaction}
\label{sec:bkg_neu}

In a deep underground environment, the neutron background is mainly created by fission and \an{} reactions as a result of natural radioactivity in the surrounding material. 
The production of neutrons by cosmic rays, whilst dominant in surface and shallow-underground laboratories \cite{Heusser1995}, is largely suppressed deep underground, thanks to the reduction of cosmic rays by the rocks overhead. 
Neutron flux measurements have been performed in different underground locations, with different detection techniques and correspondingly different capabilities to obtain information on the neutron energy spectrum \cite{Belli1989,Best2016,Bruno2019-EPJC}.
The difference in neutron flux on the Earth's surface and deep underground can be several orders of magnitude (about 3 for thermal neutrons at LNGS \cite{Csedreki2021}).

The sensitivity of a detector setup to the neutron background flux is closely related to the detection technique.
Whilst detectors based on neutron capture reactions, such as $^3$He counters, are primarily sensitive to thermalized neutrons, organic scintillators based on elastic neutron scattering on hydrogen are only sensitive to neutrons above a threshold energy.
Adding materials (mostly hydrogen rich) as neutron shielding around the detection setup may be used to alter the neutron energy spectrum (\ie{}, to thermalize the neutrons) or to reduce the neutron flux through neutron capture reactions (\eg{}~$^1$H(n,$\gamma$), $^{10}$B(n,$\alpha$), or $^6$Li(n,$\alpha$)). 

Along with the efforts to reduce backgrounds induced by environmental neutrons, backgrounds intrinsic to the detectors should also be considered as they can eventually become the limiting factor. 
In particular, $\alpha$ backgrounds may closely resemble the neutron signal. 
Material screening and selection for low intrinsic radioactivity may be required. 
Depending on the detection principle that is employed, pulse-shape discrimination may offer further improvements.
    
As an example of various of the aforementioned techniques, we describe the detection setup for a recent measurement of $^{13}$C\an{}${}^{16}$O reaction, a main source of neutrons for the s-process. 
The reaction cross section was measured at LUNA \cite{Csedreki2021, Ciani2021, BalibreaCorrea2018} using an array of eighteen stainless-steel $^3$He counters to detect the neutrons produced during the bombardment of $^{13}$C targets with a $^4$He$^+$ beam. 
The counters were embedded in a high-density polyethylene moderator to thermalize the neutrons and arranged in two concentric rings around the target chamber: the inner radius contained six counters of $25\centimeters$ active length, while the outer radius hosted twelve counters of $40\centimeters$ active length. 
This configuration allowed for a nearly 4$\pi$ solid angle coverage around the target, with an overall detection efficiency of the order of 40\% \cite{Csedreki2021}. 
The configuration of moderator and counters was chosen to optimize the sensitivity to the neutrons produced by the $^{13}$C\an{}${}^{16}$O reaction ($Q$-value $= 2.216\MeV{}$). Outer layers of borated polyethylene (with a boron content of 5\,\% by weight) were added to capture environmental neutrons and reduce their background contribution to the spectrum. 

\begin{figure}[t!]
    \includegraphics[width=\textwidth]{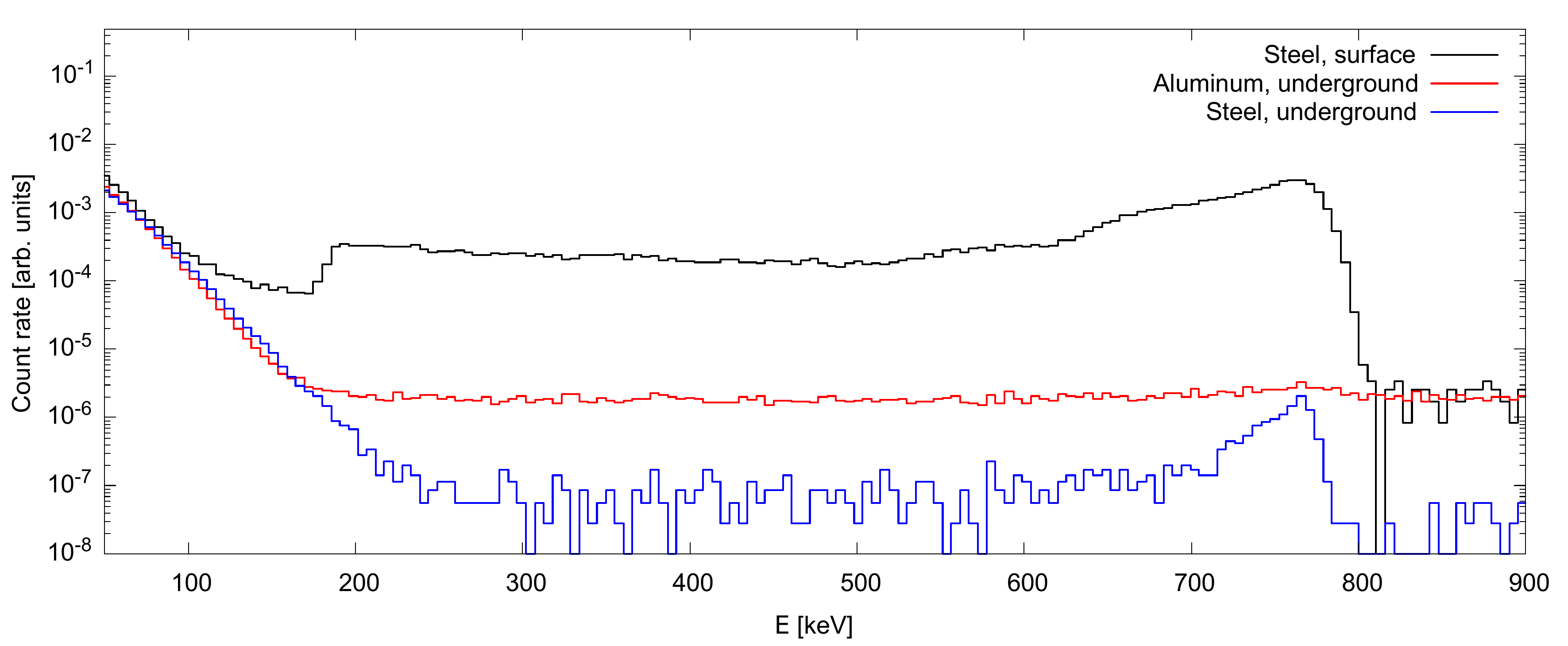}
    \includegraphics[width=\textwidth]{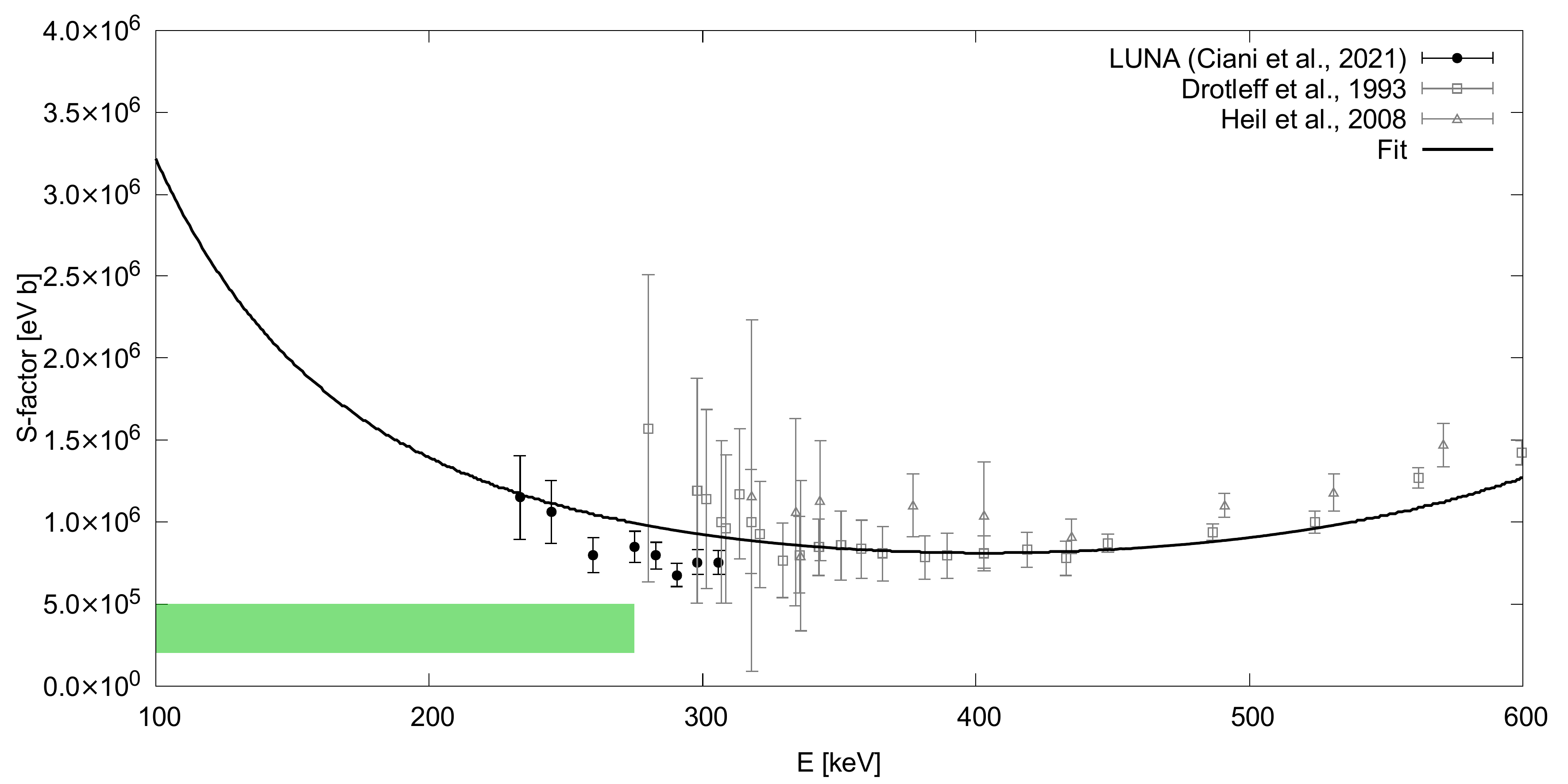}
    \caption{Upper panel: 
    Qualitative comparison between background spectra for thermalized neutrons (peak at $764\keV$) as obtained with $^3$He counters of different size  and material, overground (black) and underground (red and blue histograms).
    Note how the low levels of intrinsic radioactivity of the stainless steel housing affords a further background reduction underground compared to traditional aluminium counters.
    Lower panel: Astrophysical $S$-factor of the $^{13}$C\an{}$^{16}$O reaction.  The high precision of the LUNA data (solid symbols) \cite{Ciani2021}, for the first time within the Gamow window (green shaded area), places stronger constraints on the theoretical fit (black line). Previous literature data are taken from Refs.~\cite{Drotleff1993,Heil2008}.}
    \label{fig:13can}
\end{figure}

{\bf Figure \ref{fig:13can}} (upper panel) illustrates the effects of the underground location on the neutron background spectrum, as well as the importance of an appropriate choice of materials. 
The use of stainless steel housing provides a dramatic reduction of the intrinsic $\alpha$ background compared to prior tests using counters with an aluminium housing \cite{Best2016}. 
The two spectra acquired underground with counters made from materials of different radio-purity underlines the influence of material selection on the sensitivity of the detection setup.
In addition, pulse-shape discrimination helped to reduce background events caused by the remaining $\alpha$ activity in the walls of the detector \cite{BalibreaCorrea2018}. 

The combination of these techniques has allowed, for the first time, to reduce the background to the level required to extend the measurement of the ${}^{13}$C\an{}${}^{16}$O cross section down to the Gamow window \cite{Ciani2021}. 
The results from this measurement at LUNA are shown in comparison to literature data in the lower panel of {\bf Figure ~\ref{fig:13can}}, highlighting the role of these low energy data points in constraining a fit. 
In this energy region, the experiment is approaching a regime where the experimental yield is the limiting factor, rather than the background level. 
Measurements with high beam intensity at JUNA (see {\bf Section \ref{sec:JUNA}}) aim to overcome this limitation and reach even lower energies. 
Future measurements over a wider energy range, at JUNA with He$^{++}$ or at LUNA-MV ({\bf Section \ref{sec:LUNA MV}}) are expected to help in solving discrepancies in the normalization of different literature data sets and thus reduce the systematic uncertainty on the cross section of this important reaction.

\subsection{Activation experiments and low-background counting}
\label{sec:activation}
 
The reaction cross-section measurements discussed so far were based on the detection of the radiation ($\gamma$, charged particles or neutrons) emitted by the reaction under study. 
A completely different method to measure reaction cross sections is based on the activation of the irradiated target \cite{Gyurky2019}. 
Here the number of reactions is determined from the measurement of the radioactive decay of the reaction product. 
The technique involves two phases: first, a target is irradiated with an ion beam, and second the decay of the produced isotope is measured. 
The obvious limitation of the activation method is that the reaction product must be radioactive. 
Reactions leading to stable isotopes cannot be investigated. 
In addition, the half-life of the residual nucleus must be suitable and the decay must be followed by some kind of radiation that can be detected. 
Usually, $\gamma$ radiation is preferred as it allows the identification of the decaying isotope. 

If the basic requirements are fulfilled, the activation has some advantages compared to the conventional in-beam techniques. 
The radiation from a radioactive decay is isotropic, thus there is no need to care about angular distribution effects. 
There is no risk of missing some of the yield by non-detected weak transitions. 
The background is usually much lower because no prompt beam-induced background needs to be considered. 
Since the number of reactions is determined from the decay, the total cross section is obtained directly, which is the astrophysical quantity of interest for the reaction rate calculation. 
On the other hand, the different transitions contributing to the production of the residual nucleus are not measured, hence no spectroscopic information is gained which could be used, for example, in R-matrix fits. 
In many cases the activation can be a useful alternative approach to the cross section measurement carried out with in-beam spectroscopy, thus a comparison of the results from the two techniques can be used to identify possible systematic uncertainties. 
The combination of the two methods can lead to an improved accuracy of the final results.

In the following, we provide some examples of activation experiments in underground laboratories. 
The first group of reactions were studied by the LUNA collaboration and  both the irradiation and the decay counting was done underground at LNGS.
In a second group of reactions, the activation was performed at accelerators on surface laboratories and only the decay counting was carried out in a deep underground environment. 
Finally a case is shown where the Accelerator Mass Spectrometry was exploited. 

\subsubsection{Underground studies}

\subsubsection*{$^3$He\ag{}$^7$Be}

This reaction is one of the key reactions of the pp-chains of solar hydrogen burning \cite{Adelberger2011}. 
$^7$Be decays by electron capture to $^7$Li with a half-life of 53.22\,days and the decay is followed by the emission of a 478\keV{} $\gamma$ radiation. 
LUNA has measured the cross section of this reaction by both in-beam $\gamma$-spectroscopy and activation methods.
Prior to this measurement there was an apparent discrepancy between the results obtained with the two methods for this reaction. 
This fact largely increased the uncertainty of the cross section which made this reaction one of the least known reactions of hydrogen burning. 
The high-precision LUNA results obtained with the two methods were in excellent agreement and significantly reduced the overall uncertainty budget of this reaction \cite{Bemmerer2006-PRL,Gyurky2007,Confortola2007}. 

\begin{marginnote}
	\entry{pp-chain}{A sequence of reactions converting four protons into a $^4$He nucleus following the $\beta$ decay of two protons into neutrons with the emission of two neutrinos. 
The pp-chain represents the main hydrogen-burning mode in stars with $M \leq 1.3~M_\odot$.}
\end{marginnote}

\subsubsection*{$^{17}$O\pg{}$^{18}$F} 

This reaction takes place in explosive hydrogen burning processes like, \eg{}, in classical novae. 
The half-life of the reaction product (109.77\minutes{}) is much shorter than that of $^7$Be but still long enough so that after the irradiation the target can be removed from the reaction chamber and transported to a low background counting setup using a shielded HPGe detector.
$^{18}$F decays by positron emission and the decay is not followed by $\gamma$-emission. 
The detection of the 511\keV{} positron annihilation radiation, however, makes the activation method possible. Similar to the case of $^3$He\ag{}$^7$Be, LUNA has measured this reaction with both in-beam and activation methods. 
The low energy non-resonant cross section as well as the strength of a narrow resonance of astrophysical importance were measured with both techniques. 
The good agreement between the two approaches increased the reliability of the results also in this case and led to precise reaction rates needed for astrophysical models \cite{Scott2012,DiLeva2014}.
\begin{marginnote}
	\entry{Classical novae}{Stellar explosions driven by the thermonuclear runaway triggered on the surface of a white dwarf accreting H-rich material from a binary less-evolved companion.}
\end{marginnote}

\subsubsection*{$^{12}$C\pg{}$^{13}$N}

The proton capture of $^{12}$C is a part of the CNO cycle of hydrogen burning and strongly affects the $^{12}$C/$^{13}$C isotopic ratio during stellar nucleosynthesis. 
Its investigation is presently in progress at LUNA. 
The short half-life of $^{13}$N (9.965\minutes{}) necessitates an activation procedure different from what is discussed above. 
After irradiation, the target is not removed from the chamber, but its activity is measured with the same detector as the one used for the in-beam $\gamma$ spectroscopy. 
Nevertheless, a higher precision and reliability of the results can be expected from a combination of both approaches. \\

\subsubsection{Overground studies}

\subsubsection*{$^{169}$Tm\ag{}$^{173}$Lu and $^{144}$Sm($\alpha,\gamma$)$^{148}$Gd}

These two reactions play a role in the astrophysical p-process which is the production mechanism of the heavy, proton rich, stable isotopes \cite{Rauscher2013}. 
The long half-life of the produced isotopes (1.37 and 71.1\,years, respectively) allows for the separation of the irradiation and decay counting phases and associated measurement sites. 
On the other hand, the long half-lives result in low produced activities which necessitates the application of an extremely low background counting setup. 
In these cases the irradiations were carried out at the cyclotron accelerator of ATOMKI in Debrecen, Hungary, and then the targets were transported to a low background detector at LNGS. 
The strongly improved detection limit of the underground facility allowed for the measurement of such low cross sections which would not have been possible in an overground laboratory \cite{Kiss2011,Somorjai1998}. 

\subsubsection*{$^{25}$Mg\pg{}$^{26}$Al}

Although not an activation technique, the AMS (Accelerator Mass Spectrometry \cite{Kutschera2016}) method must also be mentioned in this section as it is also based on the direct determination of the reaction products and it also necessitates the reaction product to be radioactive. 
In an AMS experiment the number of produced isotopes is determined directly, not through their decay. 
Such an experiment does not require an underground location. 
However, it can be combined with in-beam $\gamma$ spectroscopy performed underground. 
Such an experiment has been carried out by the LUNA collaboration for the study of the $^{25}$Mg\pg{}$^{26}$Al \cite{Limata2010} which plays a role in the Mg-Al cycle of advanced H-burning. 
The irradiation and in-beam $\gamma$ spectroscopy were performed at the LUNA-400~kV accelerator, while the AMS counting of the produced $^{26}$Al isotopes was performed at the CIRCE AMS laboratory in Caserta, Italy. From the combination of the two methods high precision resonance strength values could be determined \cite{Strieder2012,Straniero2013}.


\section{THE LUNA LEGACY: 
30 Years of nuclear astrophysics underground}
\label{sec:luna-overview}

The LUNA collaboration has been pioneering underground nuclear astrophysics studies at the lowest energy frontiers for almost three decades. 
This work has led to unprecedented precision in the measurement of key reaction cross sections and to major breakthroughs in our understanding of the inner workings of stars. 
In the following, we present some key highlights from all the reactions studied so far.
A full list, together with related astrophysical scenarios and experimental approaches is given in {\bf Table \ref{tab:past}}. \\

\begin{table}[t!]\footnotesize
\begin{center}
    \caption{Overview of all reactions studied at LUNA to date. Energies are in the centre-of-mass system.}
    \begin{tabular}{| c | c | c | c | c | c |}
    \hline
    	\textbf{Reaction} & \textbf{Burning network} & {\bf Energy range [keV]} & \textbf{Target type} & {\bf Detector/Method} & \textbf{References} \\
    \hline
    \hline
    $^2$H\pg{}$^3$He & pp-chain/BBN & $2.5 - 22$ & gas &  4$\pi$-BGO & \cite{Casella2002} \\
                            & BBN & $32 - 263$ & gas & HPGe & \cite{Mossa2020-EPJA,Mossa2020} \\
    $^2$H($^3$He,p)$^4$He & pp / $e^-$ screening & $4.2 - 13.8$ & gas & Si & \cite{Prati1994,Costantini2000,Zavatarelli2001} \\
    $^2$H\ag{}$^6$Li & BBN & $80 - 133$ & gas &  HPGe & \cite{Anders2014,Trezzi2017} \\

    $^3$He($^3$He,2p)$^4$He & pp-chain & $16.5 - 24.4$ & gas & Si & \cite{Arpesella1996,Junker1998,Bonetti1999} \\
    $^3$He\ag{}$^7$Be & pp-chain/BBN & $93 - 170$ & gas &  HPGe/activation & \cite{Bemmerer2006-PRL,Gyurky2007,Confortola2007,Costantini2008} \\
    $^{6}$Li\pg{}$^{7}$Be & BBN, pre-main sequence & $60 - 350$ & solid & HPGe & \cite{Piatti2020} \\

    $^{14}$N\pg{}$^{15}$O & CNO & $70 - 228$ & gas &  4$\pi$-BGO & \cite{Bemmerer2006-NPA,Lemut2006} \\
     &  & $119 - 370$ & solid &  HPGe & \cite{Formicola2004,Imbriani2004,Imbriani2005,Marta2008,Marta2011} \\
    $^{15}$N\pg{}$^{16}$O & CNO & $90 - 230$ & gas &  4$\pi$-BGO & \cite{Bemmerer2009} \\
     &  & $70 - 375$ & solid &  HPGe/4$\pi$-BGO & \cite{LeBlanc2010-PRC,Caciolli2011-AA} \\
    $^{17}$O\pg{}$^{18}$F & CNO & $167 - 370$ & solid &  HPGe/activation  & \cite{Scott2012,DiLeva2014} \\
    $^{17}$O\pa{}$^{14}$N & CNO & $E_\mathrm{R} = 64.5$, 183 & solid &  Si & \cite{Bruno2015,Bruno2016,Lugaro2017,Straniero2017} \\
    $^{18}$O\pg{}$^{19}$F & CNO & $85 - 150$ & solid & HPGe/4$\pi$-BGO & \cite{Best2019,Pantaleo2021} \\
    $^{18}$O\pa{}$^{15}$N & CNO & $55 - 340$ & solid &  Si & \cite{Bruno2015, Bruno2019} \\
    $^{22}$Ne\pg{}$^{23}$Na & Ne-Na & $68 - 300$ & gas &  HPGe/4$\pi$-BGO & \cite{Cavanna2014,Cavanna2015,Depalo2016,Slemer2016, Ferraro2018-PRL, Ferraro2018-EPJ} \\
    $^{22}$Ne\ag{}$^{26}$Mg & s-process & $E_{\rm R} = 334$ & gas & 4$\pi$-BGO & \cite{Piatti2021}\\
    $^{23}$Na\pg{}$^{24}$Mg & Ne-Na & $E_\mathrm{R} = 138$, 240, 296 & solid & HPGe/4$\pi$-BGO & \cite{Boeltzig2019} \\
    $^{24}$Mg\pg{}$^{25}$Al & Mg-Al & $E_\mathrm{R} = 214$ & solid & HPGe/4$\pi$-BGO & \cite{Limata2010} \\
    $^{25}$Mg\pg{}$^{26}$Al & Mg-Al & $E_\mathrm{R} = 92$, 130, 189, 304 & solid & HPGe/4$\pi$-BGO & \cite{Limata2010,Strieder2012,Straniero2013} \\
    $^{26}$Mg\pg{}$^{27}$Al & Mg-Al & $E_\mathrm{R} = 326$ & solid & HPGe/4$\pi$-BGO & \cite{Limata2010} \\
    $^{13}$C\an{}${}^{16}$O & s-process & $230 - 300$ & solid & $^{3}$He counter & \cite{Ciani2021}\\   
       \hline
    $^{12}$C\pg{}$^{13}$N & CNO & $74-370$ & solid & HPGe/activation & {\rm  ongoing} \\
    $^{13}$C\pg{}$^{14}$N & CNO & $74-371$ & solid & HPGe/4$\pi$-BGO & {\rm  ongoing} \\
   $^{20}$Ne\pg{}$^{21}$Na & Ne-Na & $E_{\rm R} = 366 keV$ & gas & HPGe & {\rm ongoing} \\
    \hline
    \end{tabular}
\label{tab:past}
\end{center}
\end{table}

\noindent{\bf Big Bang Nucleosynthesis\\}
The Big Bang Nucleosynthesis era has been explored at LUNA with two different experiments carried out with the LUNA-400~kV accelerator: $^2$H\pg{}$^3$He, $^2$H\ag{}$^6$Li. 
The results on the $^2$H\pg{}$^3$He reaction (discussed in more details in section \ref{sec:bkg_gammas}) allowed to obtain the most precise determination of the primordial deuterium abundance derived so far using BBN models \cite{Mossa2020}. 
The $^2$H\ag{}$^6$Li cross section, instead, was related to the primordial $^6$Li problem. 
Astronomical observations of primordial $^6$Li in old stars find an abundance which is three orders of magnitude higher than expected according to BBN models. 
To solve this issue, it was suggested that the $^2$H\ag{}$^6$Li (producing $^6$Li) could have a resonance at low energies, increasing the cross section by orders of magnitude. 
At LUNA, the $^2$H\ag{}$^6$Li cross section was measured directly at BBN energies for the first time \cite{Anders2014,Trezzi2017}. 
Since no resonance was found, such reaction was ruled out as a possible solution to the $^6$Li problem. 
It should be noted, however, that other - non-nuclear - solutions to the $^6$Li problem also exist, see e.g. \cite{Lind2013,Steffen2012}.
Another nuclear reaction potentially related to the primordial $^6$Li problem, but linked also to pre-main sequence hydrogen burning, is the $^{6}$Li\pg{}$^{7}$Be. 
Prior to the LUNA measurement, its cross section was thought to have a broad resonance at $E_{\rm cm} = 195\keV$. 
Its existence, however, was ruled out by a recent experiment at LUNA \cite{Piatti2020}.\\

\noindent{\bf p-p chain\\}
Several experimental campaigns performed with the $50\kV$ and the $400\kV$ accelerators shed light on a number of issues related to hydrogen burning in the Sun and other stars. 
The very first experiment carried out at LUNA was the measurement of the $^3$He($^3$He,2p)$^4$He cross section, which is part of the pp chain. 
This cross section had never been measured within the solar Gamow window, and it was thought it could provide a solution to the Solar neutrino problem.
Indeed, a speculative resonance in the solar Gamow energy region could shift the balance of the pp chain towards the first branch, where no high energy neutrinos are emitted, and therefore reconcile predictions of the solar neutrino flux with observations. 
\begin{marginnote}
	\entry{Solar Neutrino Problem}{Referred to a factor of 3 discrepancy between the predicted and measured values of solar neutrinos flux on Earth. 
The puzzle was finally solved by the discovery of neutrinos oscillations.}
\end{marginnote}
At LUNA, the $^3$He($^3$He,2p)$^4$He cross section was measured directly at solar Gamow energies finding no evidence for resonances \cite{Arpesella1996,Junker1998,Bonetti1999}. 
This experiment proved for the first time the potential of underground laboratories in nuclear astrophysics. 
The following experiments at the $50\kV$ accelerator  were still focused on solar physics ($^2$H\pg{}$^3$He \cite{Casella2002}) and on understanding the electron screening effect at low energies ($^2$H($^3$He,p)$^4$He \cite{Prati1994,Costantini2000,Zavatarelli2001}). 

The installation of the LUNA-400~kV accelerator in 2001 offered a sea of possibilities to explore reactions involving heavier nuclei and energy ranges of interest for stars more massive or more evolved than the Sun. The study of the pp chain was continued with the measurement of the $^3$He\ag{}$^7$Be cross section \cite{Bemmerer2006-PRL,Gyurky2007,Confortola2007,Costantini2008}, required for a precise determination of the $^7$Be neutrino flux from solar models.\\ 

\vspace{0.5cm}
\noindent{\bf CNO cycle\\}
Since 2001, the LUNA Collaboration has been investigating most of the reactions occurring in the CNO cycle. These studies begun with the $^{14}$N\pg{}$^{15}$O reaction, the slowest of the CNO cycle. 
Multiple experiments were performed  \cite{Bemmerer2006-NPA,Lemut2006,Formicola2004,Imbriani2005,Marta2008,Marta2011}, targeting the Gamow window for shell hydrogen burning in AGB stars with unprecedented precision and improving cross-section extrapolations into the solar Gamow window. 
With our improved LUNA data, the astrophysical reaction rate at temperatures below 100\Megakelvin{} was found to be 40\% lower than the previously-adopted literature value. 
Therefore, the expected CNO solar neutrino flux, as well as the nucleosynthesis calculations for AGB stars and classical novae, were revised. 
\begin{marginnote}
	\entry{CNO cycle}{Sequence of reactions in which C, N and O nuclei act as catalysts to convert hydrogen into helium. 
The CNO cycle represents the main hydrogen-burning mode in stars with $M\geq 1.3~M_\odot$.}
\end{marginnote}As a remarkable consequence, the age of galactic globular clusters, and by extension
of the Universe, were increased by $\range{0.7}{1.0}\Gigayears$ \cite{Imbriani2004}. 

Following the success of the $^{14}$N\pg{}$^{15}$O experiments, many other crucial reactions of the CNO cycle were also studied at LUNA: namely, the $^{15}$N\pg{}$^{16}$O reaction, which initiates the NO loop of the cycle \cite{Bemmerer2009,LeBlanc2010-PRC,Caciolli2011-AA}, and the \pg{} and \pa{} branches on $^{17,18}$O. 
The ratio of the $^{17}$O\pg$^{18}$F and $^{17}$O\pa$^{14}$N cross sections determines the balance between the second and third CNO cycles and affects the isotopic abundances of oxygen and fluorine in AGB stars and classical novae \cite{Straniero2017}. 

In the $^{17}$O\pg{}$^{18}$F experiment, the reaction cross section was measured down to a centre-of-mass-energy of 167 keV, accessing for the first time the Gamow window for classical novae explosions. 
As mentioned in {\bf Section \ref{sec:activation}}, this reaction was studied by both prompt $\gamma$-ray detection and activation technique \cite{Scott2012}. 
Thanks to the consistent results obtained at LUNA, the uncertainty on the astrophysical reaction rate was reduced to 10\%, a factor of 4 smaller than adopted in previous literature. 
This placed stronger constraints on the predicted abundances of key isotopes for novae nucleosynthesis, such as $^{18}$F, $^{18}$O, $^{19}$F, and $^{15}$N \cite{DiLeva2014}. 

The study of the other reaction channel, $^{17}$O\pa{}$^{14}$N, allowed for a new determination of the strength of the $E_p = 70\keV$ resonance, which resulted in a factor-of-two increase in the astrophysical reaction rate at typical temperatures for intermediate-mass AGB stars \cite{Bruno2016} (see {\bf Section \ref{sec:bkg_part}}). 
The new proton-capture rate of $^{17}$O led to predicted $^{17}$O/$^{16}$O isotopic ratios in better agreement with those observed in Group II pre-solar grains, whose formation site was previously unknown.
This solved a longstanding puzzle on the origin of these grains, revealing how stars of 4–8 solar masses can be a likely site for their production \cite{Lugaro2017}.\\

\noindent{\bf NeNa and MgAl cycles\\}
In addition to the CNO cycle, a detailed exploration of higher-temperature hydrogen burning through the NeNa and MgAl cycles was initiated at the LUNA~400~kV accelerator, investigating directly the relevant energies of RGB and AGB stars as well as classical nova explosions. 
In particular, among the reactions of the NeNa cycle, proton captures on $^{22}$Ne \cite{Cavanna2015,Ferraro2018-PRL} and $^{23}$Na \cite{Boeltzig2019} have already been investigated, while the proton capture on $^{20}$Ne is presently under study. 
Among all milestone results, it is worth mentioning those on the $^{22}$Ne\pg{}$^{23}$Na reaction, which used to be the most uncertain reaction of the NeNa cycle, with an uncertainty of up to three orders of magnitude on the astrophysical reaction rate. 
Several resonances contribute to the $^{22}\mathrm{Ne} + \mathrm{p}$ cross section, but none of the resonances below $400\keV$ had ever been observed in direct experiments and only upper limits existed for their strengths \cite{Iliadis2010}. 
\begin{marginnote}
	\entry{NeNa and MgAl cycles}{Additional cycles for fusing hydrogen into helium, important for the synthesis of elements between $^{20}$Ne and $^{27}$Al synthesis of elements. Their energy contribution is typically negligible.}
\end{marginnote}
In addition, the possible existence of three resonances at 71, 105 and $215\keV$ had only been tentatively reported by one experiment \cite{Powers1971} but never observed in subsequent investigations \cite{Hale2001}. 
Two experimental campaigns were performed at LUNA \cite{Depalo2016,Cavanna2015,Ferraro2018-PRL}, leading to the first observation of three resonances and their related $\gamma$-decay scheme. This study placed more stringent upper limits on the unobserved tentative resonances and allowed to measure the direct capture contribution to the cross section down to unprecedented low energies. 
Thanks to the LUNA data, the new astrophysical reaction rate at AGB temperatures is now a factor of 10 higher than previously adopted \cite{Cavanna2015}, while its uncertainty has been reduced by at least two orders of magnitude.

Proton capture on $^{24,25,26}$Mg was also studied at LUNA \cite{Limata2010}. 
Efforts were especially focused on the $^{25}$Mg\pg{}$^{26}$Al reaction, which produces radioactive $^{26}$Al either in its ground state or in an isomeric state at 228 keV. 
While the isomeric state decays instantly to $^{26}$Mg, the ground state decays with a half-life of $7 \times 10^5\years$ emitting a $1.809\MeV$ gamma-ray, a signature of recent nucleosynthesis in our Galaxy. 
Two low-energy resonances of the $^{25}$Mg\pg{}$^{26}$Al reaction have been measured with unprecedented sensitivity \cite{Strieder2012}. 
As a result, the new total reaction rate is about a factor of two higher than previously suggested and the production rate of the isomeric state is up to a factor of five larger, with important consequences on the expected production sites of $^{26}$Al \cite{Straniero2013}. \\

\noindent{\bf s-process nucleosynthesis\\}
Over the last few years, investigation of nuclear reactions involved in s-process nucleosynthesis has started at the LUNA-400~kV accelerator and will continue at the future LUNA MV machine (see {\bf Section \ref{sec:LUNA MV}}). 
Two experiments have been performed so far: $^{13}$C\an{}$^{14}$N and $^{22}$Ne\ag{}$^{26}$Mg. 
The first reaction, already presented in {\bf Section \ref{sec:bkg_neu}}, is one of the main neutron sources for the s-process, and LUNA provided direct cross section data within the Gamow window for the first time \cite{Ciani2021}.
The latter is the only open channel in $^{22}$Ne+$\alpha$ fusion at low energies ($E < 565$~keV), while at higher energies the \ag{} channel competes with the $^{22}$Ne\an{}$^{25}$Mg neutron source.
The nucleosynthesis of isotopes between $^{26}$Mg and $^{31}$P in massive AGB stars is affected by the uncertainty of the $^{22}$Ne\ag{}$^{26}$Mg reaction rate \cite{Karakas2006}.
At temperatures below 0.3 GK the $^{22}$Ne\ag{}$^{26}$Mg reaction rate is dominated by a weak resonance at 334 keV in the centre of mass. 
Such resonance also affects the cross-over temperature where the \an{} rate starts to exceed the \ag{} rate \cite{Karakas2006}. 
The 334 keV resonance was tackled at LUNA using the gas target system and the 4$\pi$-BGO detector \cite{Piatti2021}.
\begin{marginnote}
	\entry{s-process nucleosynthesis}{Refers to a sequence of neutron-capture reactions that are {\em slow} (hence the name) compared to the $\beta$ decays of unstable nuclei produced along the path. The process evolves close to the valley of $\beta$ stability and contributes to the synthesis of elements heavier than iron.}
\end{marginnote}

\subsection{Future opportunities: LUNA MV}
\label{sec:LUNA MV} 

Despite the impressive progress of the last decades, many other important reactions remain beyond the reach of the existing LUNA-400~kV accelerator, partly because of the limitation to a maximum beam energy of $400\keV$. 
This energy is typically enough for the study of most of the hydrogen burning reactions in or close to their Gamow windows. 
However, in those cases where the Gamow window cannot be reached,  cross section measurements are needed over as wide an energy range as possible to aid theoretical extrapolations. 
Also, reactions of more advanced burning processes, such as for example helium- and carbon burning, require even higher beam energies as they take place at correspondingly higher temperatures. 

The need for a higher energy accelerator deep underground at LNGS  was formulated by the LUNA collaboration several years ago and also endorsed by NuPECC \cite{NUPECC}. 
The assessment of the technical requirements of the new accelerator as well as the careful selection of its location at LNGS have already been carried out and the year 2021 finally witnessed the installation of the new LUNA MV machine. 
This project was funded by the Italian Ministry of Education, University and Research.

The LUNA MV accelerator is a single-ended Inline Cockcroft Walton accelerator with a maximum $3.5\MV$ terminal voltage. 
It was constructed by the High Voltage Engineering Europe and optimized for high long- and short-term energy stability, long duty cycle and long term operation without personnel on site \cite{Sen2019}. It was installed at the north side of Hall B in the LNGS laboratory in a newly constructed dedicated building which provides the necessary radiation shielding for the rest of the underground lab. 
The accelerator is able to deliver proton-, alpha- and $^{12}$C beams in the $\range{300\keV}{3.5\MeV}$ energy range with beam intensities up to 1000, 500 and $150\emuA$, respectively, for the three ion species. 
By selecting the $^{12}$C$^{++}$ charge state, the energy range for carbon beam can be extended to $7\MeV$. 

The LUNA collaboration has formulated a rich scientific program for the new accelerator in the forthcoming years. 
The first reaction to be studied will be the $^{14}$N\pg{}$^{15}$O, for which the collaboration already has extensive experience in its low energy study at the $400\kV$ accelerator. 
Higher energy cross section measurements, on the other hand, are needed as the extrapolated cross section to solar energies is not precise enough for astrophysical models \cite{Adelberger2011}.
These facts make the $^{14}$N\pg{}$^{15}$O reaction an ideal pilot project of LUNA MV. 
The cross section will be measured from the $E_\mathrm{p} = 278\keV$ resonance up to the maximum energy of the accelerator providing a high precision data set over a wide energy range overlapping with the available data from the LUNA-400~kV accelerator.

The availability of carbon beam at LUNA MV will enable the study of the $^{12}$C+$^{12}$C reaction, arguably one of the most important in nuclear astrophysics as it regulates the energy generation and nucleosynthesis during the carbon burning phase of  massive stars \cite{Beck2020} and dictates whether they will explode as supernovae. In the literature, low quality and often contradicting cross section data can be found which, moreover, do not reach low enough energies \cite{Tan2020}. 
At the LUNA MV accelerator the proton and alpha channels of the reaction will be measured -- supplemented by the detection of $\gamma$ rays from the excited states of residual nuclei -- at energies lower than ever before, including the search for possible low energy resonances. 

A natural continuation of the $^{13}$C\an{}${}^{16}$O cross section measurement already performed at the LUNA-400~kV accelerator is the higher energy study of both s-process neutron-source reactions: $^{13}$C\an{}${}^{16}$O and $^{22}$Ne\an{}$^{25}$Mg. 
The complicated structure of the $^{13}$C\an{}${}^{16}$O excitation function and the uncertain overall normalization requires cross-section measurements over a broad energy range. 
This will be achieved with a setup similar to the one used at LUNA-400~kV. 
The possibility of using inverse kinematics is also considered, which would likely afford conditions of reduced beam-induced background. 
The $^{22}$Ne\an{}$^{25}$Mg reaction is the main source of neutrons for the s-process in massive stars. 
For its study, a novel type of neutron detector will be developed as part of an ERC project, SHADES (Scintillator-$^3$He Array for Deep-underground Experiments on the s-process). 
The new detector array, combined with the ultra low background environment and the high beam intensity of LUNA MV, will allow to measure high accuracy cross sections about two orders of magnitude lower than before for this reaction. Exploring the whole Gamow window will drastically reduce the uncertainty of the astrophysical reaction rate in the relevant temperature range.


\section{OTHER UNDERGROUND LABORATORIES WORLDWIDE}
\label{other-labs}

Not surprisingly, the success story demonstrated by the pioneering work at LUNA has prompted worldwide efforts for the installation of similar accelerators in other underground laboratories. 
Some of these initiatives will be briefly presented in the next section.

\subsection{CASPAR, US}

CASPAR, the Compact Accelerator System for Performing
Astrophysical Research, was the first accelerator to be commissioned at an underground site in the United States. Located at the Sanford Underground Research Facility (SURF) in the former
Homestake Gold mine (Lead, South Dakota) it was designed for nuclear astrophysics measurements at a depth of 4850\,ft ($1478\meters$), affording a shielding against cosmic radiation of about 4300\meters{}-water-equivalent \cite{Heise2020}. 
The $1\MV$ Van~de~Graaff accelerator already had a fruitful history in research starting long before CASPAR \cite{Wiescher2017}. 
After upgrades to increase the beam intensity and modernize the control system, the accelerator was installed at SURF, where first beam was produced in 2017 \cite{Heise2020}.

With an operational range of $150\kV$ to $1100\kV$ terminal voltage \cite{Aliotta2021} and beam intensities of up to $250\muA$, CASPAR extended the available energy range for proton and helium beams at underground facilities, previously set by LUNA-400~kV, and thus opened up opportunities for new underground studies. 
With the availability of a solid target and a windowless gas target setup, a wide range of reactions can now be studied at CASPAR. 
As a commissioning experiment, $^{14}$N\pg{}$^{15}$O was chosen and studied using solid targets.
HPGe or high-efficiency NaI(Tl) detection setups are available for the study of radiative capture reactions, complemented by an array of ${}^3$He counters for neutron detection. 
More recent measurements include ${}^{10}\mathrm{B}(\alpha,\mathrm{n}){}^{14}\mathrm{N}$ and ${}^{7}\mathrm{Li}(\alpha,\gamma){}^{11}\mathrm{B}$ \cite{Aliotta2021} as well as 
 ${}^{22}$Ne\an${}^{26}$Mg \cite{Strieder2019}. 
Data analysis is currently in progress. 
A rich scientific program is planned for the study of key reactions at CASPAR, complementing the capabilities of the other underground accelerator facilities.

\subsection{JUNA, China}
\label{sec:JUNA}

The Jinping Underground Nuclear Astrophysics (JUNA) laboratory is located in the Sichuan province, China, under the Jinping mountain which provides a shielding of 2400\meters{} (7620\,\mwe{}) of radioactively inert marble rocks and affords a cosmic-ray background 2 orders of magnitude lower than at Gran Sasso \cite{Liu2016}.
A 400\kV{} accelerator, coupled to a $2.45\GHz$  Electron Cyclotron Resonance (ECR) source, is capable to deliver up to 10\emA{} proton- and  6\emA{} $^4$He$^+$ beams \cite{Liu2016}. 
Space charge effects and beam transport efficiency are optimized by the use of a Low Energy Beam Transport Line (LEBT).
The accelerator started operation in December 2020 and four experiments have already been performed \cite{Aliotta2021}, as briefly summarized below. 

The $^{12}$C($\alpha,\gamma$)$^{16}$O reaction, which regulates the C/O abundance at the end of helium burning and affects the subsequent phases of stellar evolution, was studied in direct kinematics using a 1\emA{} $^4$He$^+$ beam onto a pure $^{12}$C target surrounded by an array of BGO and LaBr detectors; the reaction cross section was measured down to $E_{\rm cm} = 600\keV{}$, the lowest to date.
The $^{13}$C\an{}${}^{16}$O reaction, the main source of neutrons for the s-process in Asymptotic Giant Branch stars at temperatures $T=90\Megakelvin$, was studied down to $E_{\rm cm} = \range{400}{600}\keV{}$ using intense ($\range{0.1}{2}\pmA{}$) beams of $^4$He$^{1+}$ and $^4$He$^{2+}$ on $2\millimeters$ thick $^{13}$C targets; neutrons were detected with an array of 24 $^3$He counters arranged in concentric rings.
Measurements of the $^{25}$Mg\pg{}$^{26}$Al reaction, important for the synthesis of $^{26}$Al in the Galaxy, have focused on the width of two important resonances at 92 and 189\keV{}. Thick target yield measurements were performed using a 4$\pi$ BGO $\gamma$-ray detector. 
Finally, the $^{19}$F(p,$\alpha \gamma$)$^{16}$O and $^{19}$F\pg{}$^{20}$Ne reactions, important in the CNO cycles, were studied in a combination of overground and underground measurements, these latter extending down to $E_{\rm cm} = 72\keV{}$ and $188\keV{}$ for the two reactions respectively. 
A new resonance, observed at $225\keV{}$ in $^{19}$F\pg{}$^{20}$Ne, enhances the rate by a factor of~4, thus increasing leakage from the CNO cycle and possibly explaining Ca abundance in the first generation Pop III stars \cite{Aliotta2021}. 
Results from the $^{19}$F(p,$\alpha \gamma$)$^{16}$O study have just been published \cite{Zhang2021}.
The scientific program at JUNA is expected to continue with the potential for major breakthroughs in nuclear astrophysics research.

\subsection{Felsenkeller, Germany} 

Unlike the deep underground laboratories presented above, the Felsenkeller ion accelerator laboratory in Dresden, Germany, is a shallow underground site \cite{Felsenkeller}. 
It is located under 45\,meters of hornblende monzonite rock overburden (140\,meters water equivalent). Such a depth is enough to completely shield all the components of cosmic-ray induced radiations except muons. 
The significant remaining muon flux necessitated a detailed study of background conditions of the site and the optimization of detection techniques. 
Dedicated experiments were devoted to assess $\gamma$-ray \cite{Szucs2019}, neutron- \cite{Grieger2020} and muon backgrounds \cite{Ludwig2019}. 
An important conclusion was that the combination of the shallow underground location with active detector shielding results in a background rate in $\gamma$ detectors
which is only a factor $2-3$ worse than in a deep underground laboratory at $\gamma$-ray energies of $\range{5}{8}\MeV$ \cite{Szucs2012}. 
Such a background is low enough for nuclear astrophysics studies and a shallow underground accelerator can be a good alternative or a complementary facility to deep underground ones. 
Encouraged by these results, a 5\,MV Pelletron tandem accelerator has recently been installed in the Felsenkeller laboratory. 
Thanks to the additional ion source on the high voltage terminal, the accelerator can be used both in tandem and single-ended modes and can deliver proton, $\alpha$ particle and $^{12}$C ion beams with several tens of $\muA$ intensity.

The scientific program of the Felsenkeller underground ion accelerator laboratory \cite{Bemmerer2019} foresees the study of several reactions of astrophysical importance such as $^3$He\ag{}$^7$Be and $^{12}$C\ag{}$^{16}$O. 
To fully exploit the background reduction capabilities, a number of HPGe detectors surrounded by active veto detectors will be used for the experiments.


\section{CONCLUSIONS AND OUTLOOK}

The last few decades have witnessed a tremendous development in astronomy and astrophysics. 
Astronomical observations have become astoundingly precise and supercomputers now allow detailed 3D modelling of stellar interiors at various stages of stellar evolution. 
Consequently, we have a better understanding of astrophysical phenomena and a deeper knowledge of the elemental composition of the universe. 
On the other hand, the improved description of microscopic processes in stars is also needed. 
Nuclear reactions generate the energy of stars and produce the chemical elements, so knowing the characteristics of these reactions is necessary in modern astrophysics. 

Since the birth of nuclear astrophysics in the middle of the 20$^{\rm th}$ century, 
many fusion reactions between stable nuclei have been studied experimentally and have led to a broad understanding of stellar evolution and nucleosynthesis.
However, the extremely low cross sections encountered at stellar temperatures prevented for a long time the study of reactions directly at the relevant energies. 
Cutting-edge experimental techniques are needed for such studies and require  intense ion beams, high efficiency detection systems and modern nuclear electronics. 
Background radiations from environmental radioactivity or of cosmic origin may still hinder the measurement of low cross sections. 
The effective reduction of the background is thus critically important. 

The most effective suppression of cosmic-ray induced background can be achieved by placing the experiment in a deep underground laboratory shielded by hundreds to thousands meters of rock. 
In the last 30 years, the LUNA (Laboratory for Underground Nuclear Astrophysics) collaboration -- operating for a long time the world's only deep underground accelerator -- proved that a deep location combined with advanced experimental techniques can lead to measured cross sections much lower and much more precise than previously achieved. 

In this paper we have summarized the experimental requirements and solutions for a successful measurement of low reaction cross sections in an underground laboratory. 
As examples, the methods used by the LUNA collaboration were shown, and highlights of key scientific achievements presented. 
However, the fast development of observational and theoretical astrophysics cited above necessitates further progress in experimental nuclear astrophysics. 
There are many cases where nuclear cross sections represent the major source of uncertainty in stellar models. 
The unique conditions offered by an underground accelerator need to be  further exploited to improve the nuclear physics input of astrophysical calculations. 
Besides the imminent upgrade of the LUNA project with a higher energy accelerator, other worldwide initiatives for deep underground nuclear astrophysics experiments have recently become operational. 
These facilities will help nuclear physics to keep pace with high precision observations in astrophysics.

\begin{summary}[SUMMARY POINTS]
\begin{enumerate}
	\item Thermonuclear reactions in stars take place over a narrow energy region commonly referred to as the Gamow peak. The peak curve arises from the product of the Maxwell-Boltzmann distribution at a given stellar temperature $T$ and the quantum probability of tunnelling through the Coulomb barrier between the interacting nuclei.
	\item The cross sections of thermonuclear reactions drop exponentially at Gamow energies and are extremely challenging to study in terrestrial laboratories.
	\item Underground laboratories, with their significantly reduced cosmic-ray backgrounds, provide an ideal environment to study nuclear reactions of astrophysical interest.
	\item Major progress has been achieved at the Laboratory for Underground Nuclear Astrophysics (LUNA) located under the Gran Sasso mountain in Italy. The 1.4\unit{km} rocks overburden affords significantly reduced backgrounds for the detection of $\gamma$ rays, charged particles and neutrons.
	\item Over the last 30 years, the LUNA collaboration has studied key hydrogen-burning reactions relevant to Big Bang Nucleosynthesis, the pp-chain and the CNO cycle, the NeNa and MgAl cycles.
	These studies have led to an improved understanding of energy generation and nucleosynthesis in various astrophysical sites, including the Sun, red giant and asymptotic giant branch stars, and classical novae.
\end{enumerate}
\end{summary}

\begin{issues}[FUTURE ISSUES]
\begin{enumerate}
	\item Despite the impressive progress achieved so far, many key reactions of astrophysical interest remain beyond technical capabilities at the existing LUNA-400~kV accelerator.
	\item A new 3.5\MV{} accelerator recently installed at Gran Sasso will open up unprecedented opportunities for the study of key nuclear reactions of helium- and carbon burning, including the two neutron sources, $^{13}$C\an{}${}^{16}$O and $^{22}$Ne\an{}$^{25}$Mg, and the $^{12}$C($^{12}$C,p)$^{23}$Na and $^{12}$C($^{12}$C,$\alpha$)$^{20}$Ne reactions.  A further study, $^{14}$N\pg{}$^{15}$O is also planned during the commissioning of the new accelerator and will serve to investigate  the core metallicity of the Sun.
	These studies will help shed light on more advanced stages in the evolution of massive stars.
	\item New underground laboratories dedicated to nuclear astrophysics studies have become operational in recent years, both in the US (CASPAR) and in China (JUNA). Initial scientific results have recently been published.
	\item Shallow underground laboratories, such as Felsenkeller, do not offer the same level of background suppression as deep underground laboratories and generally necessitate of active veto detectors. Nevertheless, they provide a useful complementary site for the study of astrophysical reactions for which background issues are not the limiting factor.
\end{enumerate}
\end{issues}


\section*{DISCLOSURE STATEMENT}

The authors are not aware of any affiliations, memberships, funding, or financial holdings that
might be perceived as affecting the objectivity of this review. 

\section*{ACKNOWLEDGMENTS}

This work was supported, in part, by the Science and Technology Facilities Council, UK (grant
no. STFC ST/P004008/1), and by National Research, Development and Innovation Office
(NKFIH) grant no. K134197. M.A. also acknowledges financial support from the ExtreMe Mat-
ter Institute during her Visiting Professorship at GSI/Goethe University Frankfurt. A.B. received
funding from the European Union under European Research Council (ERC) grant agreement
852016 (SHADES) and grant agreement 101008324 (ChETEC-INFRA). For the purpose of open
access, the authors have applied a Creative Commons Attribution (CC BY) license to any Author
Accepted Manuscript version arising from this submission.


\end{document}